\documentclass[preprintnumbers,amsmath,amssymb,11pt]{revtex4-1}
\usepackage{graphicx}
\usepackage{dcolumn}
\usepackage{bm}
\usepackage[colorlinks,bookmarksopen,bookmarksnumbered,citecolor=red,urlcolor=red]{hyperref}
\usepackage{filecontents}
\begin{document}

\title{Low frequency modulation of jets in quasigeostrophic turbulence}
\author
{Suhas D.L.$^1$ and Jai Sukhatme$^{1,2}$ \\
   1. Centre for Atmospheric and Oceanic Sciences, Indian Institute of Science, Bangalore 560012, India \\
   2. Divecha Centre for Climate Change, Indian Institute of Science, Bangalore 560012, India \\}
\date{\today}
\begin{abstract}

Quasigeostrophic turbulence on a $\beta$-plane with a finite deformation radius is studied numerically, with particular emphasis on
frequency and combined wavenumber-frequency domain analyses.
Under suitable conditions,
simulations with small-scale random forcing and large-scale drag exhibit a spontaneous formation of multiple
zonal jets.
The first hint of wave-like features is seen in the distribution of kinetic energy as a function of frequency; specifically,
for progressively larger deformation scales there are systematic departures in the form of isolated peaks (at progressively higher frequencies)
from a power-law scaling. Concomitantly, there is an inverse flux of kinetic energy in frequency space which extends to
lower frequencies for smaller deformation scales.
The identification of these peaks as Rossby waves is made possible by
examining the energy spectrum in frequency-zonal wavenumber and frequency-meridional wavenumber diagrams.
In fact, the modified Rhines scale turns out to be a useful measure of the dominant meridional wavenumber of the modulating Rossby waves;
once this is fixed,
apart from a spectral peak at the origin (the steady jet), almost all the energy is contained in westward propagating disturbances that follow
the theoretical Rossby dispersion relation.
Quite consistently, noting that the zonal scale of the modulating waves is restricted to the first few wavenumbers,
the energy spectrum is almost entirely contained within the corresponding Rossby dispersion curves on a frequency-meridional wavenumber diagram.
Cases when jets do not form are also considered; once again, there is a hint of Rossby wave activity, though the spectral peaks are quite muted. Further, the kinetic energy 
scaling in frequency domain follows a 
-5/3 power-law 
and is distributed much more broadly in
frequency-wavenumber diagrams.

\begin{center}
{\bf Journal ref: Physics of Fluids, 27, 016601, 2015.}
\end{center}

\end{abstract}

\maketitle

\section{Introduction}

Quasigeostrophic (QG) turbulence, or QG dynamics in the nonlinear regime, involves a subtle interplay of
turbulent and wave-like motions \citep{Rhines}. The QG equation in dimensional form reads \citep{Vallis-book},
\begin{equation}
\frac{D}{Dt}[\triangle \psi - \frac{1}{L_D^2} \psi] + \beta \frac{\partial \psi}{\partial x} = 0.
\label{1}
\end{equation}
Here $D/Dt$ is the two-dimensional (2D) material derivative, $\psi$ is the 2D streamfunction, $L_D$ is the Rossby deformation scale and
$\beta$ is the co-efficient of the linear term in the expansion of the Earth's Coriolis parameter.
Following Rhines \cite{Rhines}, QG turbulence has, and continues to be, extensively studied; in particular,
a striking feature of QG turbulence is the spontaneous emergence of jets or zonal flows from isotropic data
(see for example, \citep{HH,Maltrud-Vallis,Cheklov-etal,Marcus,Danilov1,Gal-etal,Smith-qg,SDG-2007,McDrit,Scott1}).

As it happens, much of the work on QG turbulence has focussed on the
limit $L_D \rightarrow \infty$, i.e.\ an infinite deformation scale.
When $L_D = \infty$, jet formation is anticipated due to
two constraints; namely,
the transfer of energy to larger scales, and
to smaller frequencies \citep{Rhines,Hass}.
The dispersion relation of Rossby waves (when linearized
about a state of rest) reads,
\begin{equation}
\omega(\vec{k}) = \frac{-\beta k_x}{k_x^2 + k_y^2 + 1/L_D^2}.
\label{2}
\end{equation}
Therefore, if $L_D = \infty$, it is not possible to maintain isotropy (i.e.\ $k_x \sim k_y$) while satisfying
both these constraints. An expected consequence is the ``flattening of eddies" or a transfer of energy to modes with $k_x \ll k_y$; in other words, the
generation of zonal flows \citep{Rhines,HH}. To be sure, there is more to jet formation than these constraints; for example, two (not necessarily unrelated)
viewpoints include (i) a systematic isotropic inverse transfer of energy which at some length scale is diverted to
zonal modes (see for example the elaborate discussion in Sukoriansky et al. \cite{SDG-2007}), with near-resonant interactions
playing a prominent part in this process \citep{Smith-qg} and (ii)
jet formation via a modulational \citep{Cetal}, or
quasilinear symmetry breaking instability \citep{SrinYoung} --- the reader is referred to Bakas and Ioannou \cite{BI-2014} (and the 
references therein) for insight 
into the rapidly growing literature on stochastic structural stability theory that addresses the 
emergence of zonal and other coherent structures in the QG system.
More broadly, Connaughton et al. \cite{Con} provide a useful and up-to-date review of instabilities, the role of invariants, and classes of nonlinear
interactions thought to play a role
in zonal flow
formation.

\subsection{Finite $L_D$ and $\beta \neq 0$}

Physically, $\beta$ in (\ref{1}) allows for the presence of Rossby waves \citep{Vallis-book}, while a finite $L_D$ 
reflects the effects of a
free upper surface in the QG dynamics of a
single shallow water layer of fluid \citep{Larichev-Mc}. Another way to look at (\ref{1}) is to view it as a QG approximation to the dynamics
of the active layer in a two layer system (with the other layer being inert). In this framework, the finite deformation radius is a measure of the
reciprocal of the divergence of the flow \citep{Lipps}.
Even though effects of a finite $L_D$ and $\beta \neq 0$ on the linear stability of jets and
predictability of geophysical flows
have a long history (see for example, Lipps \cite{Lipps} and Holloway \cite{Holloway}, respectively)
this regime has attracted renewed attention in the context of the formation of jets on
Jupiter and Saturn \citep{Theiss,Penny}, in the Earth's oceans \cite{Eden}, and more generally as a local model for planetary atmospheres \citep{Smith-2004}.
Indeed, as described by Smith \cite{Smith-2004}, the presence of both these terms together can have significant effects on the nonlinear evolution of
the system (interestingly, when $L_D$ is
finite, $\beta$ is usually set to zero, and studies have focussed on the so-called
equivalent barotropic model \citep{Larichev-Mc,KOY-1995,ISW-2002,Tran-Bowman,Tran-Drit,Scott2}.
In fact, the limit $L_D \rightarrow 0$ of the equivalent barotropic model, referred to as the asymptotic \citep{Larichev-Mc}, or
``ideal 2D" limit \citep{ISW-2002}, has also
received a fair amount of attention.).

For example, as is seen from (\ref{2}), with finite $L_D$, it is possible to satisfy the aforementioned constraints
of energy transfer to large scales and low frequencies
(similar features of 2D dispersive systems with different degrees of locality were pointed out in
Sukhatme and Smith \cite{SS-2009}).
Bearing this in mind, it has been suggested that anisotropy in the flow manifests itself when there is
frequency matching between the (isotropic) turbulent flow and Rossby waves \citep{Okuno,Smith-2004,Theiss}. Specifically,
the length scale ($L_T$)
at which the transition to a zonal flow occurs is given by \citep{Okuno},
\begin{equation}
L_T = 1/[\frac{1}{L_R^2} - \frac{1}{L_D^2}]^{\frac{1}{2}}, ~~~ \textrm{where}~~L_R=\sqrt{\frac{U}{\beta}}.
\label{3}
\end{equation}
As $L_D \rightarrow \infty$, $L_T \rightarrow L_R$, where $L_R$ is the Rhines scale. When $L_D$ decreases below $L_R$, (\ref{3}) gives
an unphysical result.
Arguing on these lines, Okuno and Masuda \cite{Okuno} concluded that there is an upper limit to the divergence
(lower limit to $L_D$, specifically $L_D \ge L_R$)
where zonal flows are possible.
It is interesting to note that the condition for
the marginal barotropic stability of a
jet (of a single scale $L_T$) yields the same estimate as (\ref{3}) \citep{Smith-2004}.
In the presence of a prescribed mean flow, it turns out that
for a fixed $\beta$ and $L_D$, there is a critical value of the mean flow beyond which frequencies cannot match, and one does not
expect the zonation to take place \citep{Theiss}. Quite remarkably, this argument works well for
the observed latitudinal distribution of isotropic
vortices and jets in Jupiter and Saturn \citep{Theiss,Penny} (see also, Eden \cite{Eden} for a discussion in the context of the Earth's oceans).

Examining (\ref{3}), Smith \cite{Smith-2004} noted that $L_R$ depends on $U$, and that $U$ is a part of the
problem, rather than an externally prescribed entity.
Taking the problem to be forced with a fixed energy source, and assuming an inertial range where energy is
predominantly transferred upscale, it becomes possible to provide an estimate of $U$. Viewed in this manner, the only
free parameter is $\beta$; thus,
for given spectral properties of energy and its rate of input into the system, at fixed $L_D$,
there is a threshold $\beta$ for which zonal flows are expected to form
\citep{Smith-2004}.

\subsection{Frequency domain}

Interestingly, even though the analysis of QG turbulence in wavenumber domain has a long history, the distribution, scaling and fluxes in frequency domain
have only been examined recently. For example, with an infinite deformation radius, Sukoriansky et al. \cite{SDG-2007,Suko-PRL} show that Rossby waves
co-exist with jets in
a fully developed turbulent state (see also the discussion on non-zonal coherent structures in Bakas and Ioannou \cite{BI-2014}); a fact supported by the analysis of laboratory experiments \citep{Afan}.
Notably, the comprehensive study by Arbic et al. \cite{Arbic-2012} of a two-layer $f$-plane QG system (see also, Berloff and Kamenkovich \cite{BK1,BK2}),
ocean circulation models and surface
current data has added significantly to the overall picture of energy distribution in the frequency domain.

In the present work, we focus on (\ref{1}) and consider both $L_D=\infty$ and well as finite $L_D$ in the presence of the $\beta$ effect. The formation of
jets or vortices (depending on $L_D$), scaling of spectra in wavenumber space as well as the inverse transfer of kinetic and potential energy is verified
(Section 2).
We then proceed to
examine the emergent fields in the frequency domain; a systematic transfer of energy to smaller frequencies and the establishment of a
power-law scaling is noted.
In the jet forming cases, it is seen that the inverse transfer only proceeds up to a certain frequency, 
below which the spectra show significant isolated peaks. On the other hand, the inverse transfer and power-law (close to -5/3 slope) holds over a 
larger range of frequencies
without jets.
Even though the shifting of
the peaks to lower frequencies with an increasing deformation radius is consistent with the Rossby dispersion relation;
it is the decomposition the spectra into frequency-zonal wavenumber and
frequency-meridional wavenumber diagrams that makes it possible to
identify the modulating disturbances as Rossby waves and quantify their scales (Section 3). 
Finally, the results are collected and discussed in Section 4.

\section{Numerical setup and data analysis}

For purposes of numerical simulation we prefer to deal with a non-dimensional form of (\ref{1}); in particular, choosing
length and velocity
scale parameters $L,U$, respectively, this yields,
\begin{equation}
\frac{D}{Dt}[\triangle \psi - \epsilon^2 \psi] + \frac{1}{R_h} \frac{\partial \psi}{\partial x} = 0,
\label{4}
\end{equation}
where, $\epsilon=L/L_D$ and $R_h=U/\beta L^2$ (Rhines number) are two non-dimensional parameters.
For $L_D = \infty$, the length scale that emerges naturally from (\ref{1}) is the Rhines scale $L_R$. Therefore,
choosing $L=L_R \Rightarrow R_h \sim O(1)$ while $\epsilon=L_R/L_D$. In the simulations we consider a series of deformation
scales, namely, $L_D=\infty, 1/2, 1/5, 1/7, 1/10, 1/12$ and 1/15, thus for a fixed $L_R \approx 1/5$,
we span $0 \le \epsilon \le 3$, i.e., we straddle the balance $\epsilon =1$ where $L_R=L_D$ (note that $L_R$ depends on $U$ and is
determined {\it a posteriori}, in fact, it varies between 1/4.5 and 1/6.5 in the various cases considered).

The numerical simulations are pseudo-spectral in nature,
the domain is doubly periodic, the forcing is spectrally localized at small scales, and time
stepping is done via a Runge-Kutta fourth order integrator. A high order hyperdissipation is employed
to remove energy at small scales
and linear drag is used to extract energy at large scales.
In particular, the energy sink comprises of ($\lambda+\nu\triangle^{n})\zeta$, where $\lambda$ represents the linear drag,
$\nu$ is the hyperviscosity and $\zeta=\triangle\psi-\psi/L_D^2$.
In all our simulations $\lambda=10^{-3}$ and $\nu=20/(k_{\textrm{max}})^{2n}$ where $n=4$ and $k_{\textrm{max}}=0.94 \times N/2$,
$N$ being the number of grid points \citep{Danilov2001}.
A random Markovian formulation is used for forcing ($F$), i.e.,
$F_n = \hat{A}(1-R^2)e^{i\theta}+RF_{n-1}$,
where $\theta$ is a random number in $[0,2\pi)$, $\hat{A}$ is the wavelength-dependent forcing amplitude, $R$ is the correlation coefficient
and the subscript on $F$
denotes the timestep \citep{Maltrud-Vallis}. The forcing is restricted to $(k_f-2,k_f+2)$. In our simulations $R=0.5$ and $k_f=50$.
All results are reported at a resolution of $256 \times 256$, though a variety of cases were tested at 512 $\times$ 512 
resolution.

Apart from spectra and fluxes in the wavenumber domain, we analyze the data generated in combined frequency-wavenumber diagrams. These diagrams
are widely used in atmospheric analyses, especially tropical meteorology (see for example, Wheeler and Kiladis \cite{WK-1999}).
In particular, they allow a
direct comparison with theoretically calculated dispersion relations, and thus an identification of the kind of waves
that are present in the data being examined. Specifically,
a longitude-time plot of the zonal velocity is generated by selecting a particular latitude. This data is then 
transformed and  normalized to obtain a frequency-zonal wavenumber spectra. Similarly, by fixing the longitude and generating a latitude-time plot,
frequency-meridional wavenumber plots are obtained.
The energy flux is calculated by \citep{Arbic-2012},
\begin{equation}
\Pi(K,\omega) = \int_{k^2+l^2 \geqslant K^2} \!\! \int_{\omega' \geqslant \omega} \!\!\!\!\! T(k,l,\omega') ~d\omega' dk dl ,
\label{5}
\end{equation}
where $T = \Re[\hat{J}(\psi,\triangle \psi) \hat{\psi}^*]$, and $(\cdot)^*$ denotes a complex conjugate.
Equation (\ref{5}) is averaged over all $K$ to obtain $\Pi(\omega)$, the flux as a function of frequency.

\section{Results}

\subsection{Analysis in wavenumber space}

\begin{figure}
\centering
\includegraphics[width=7cm,height=6cm]{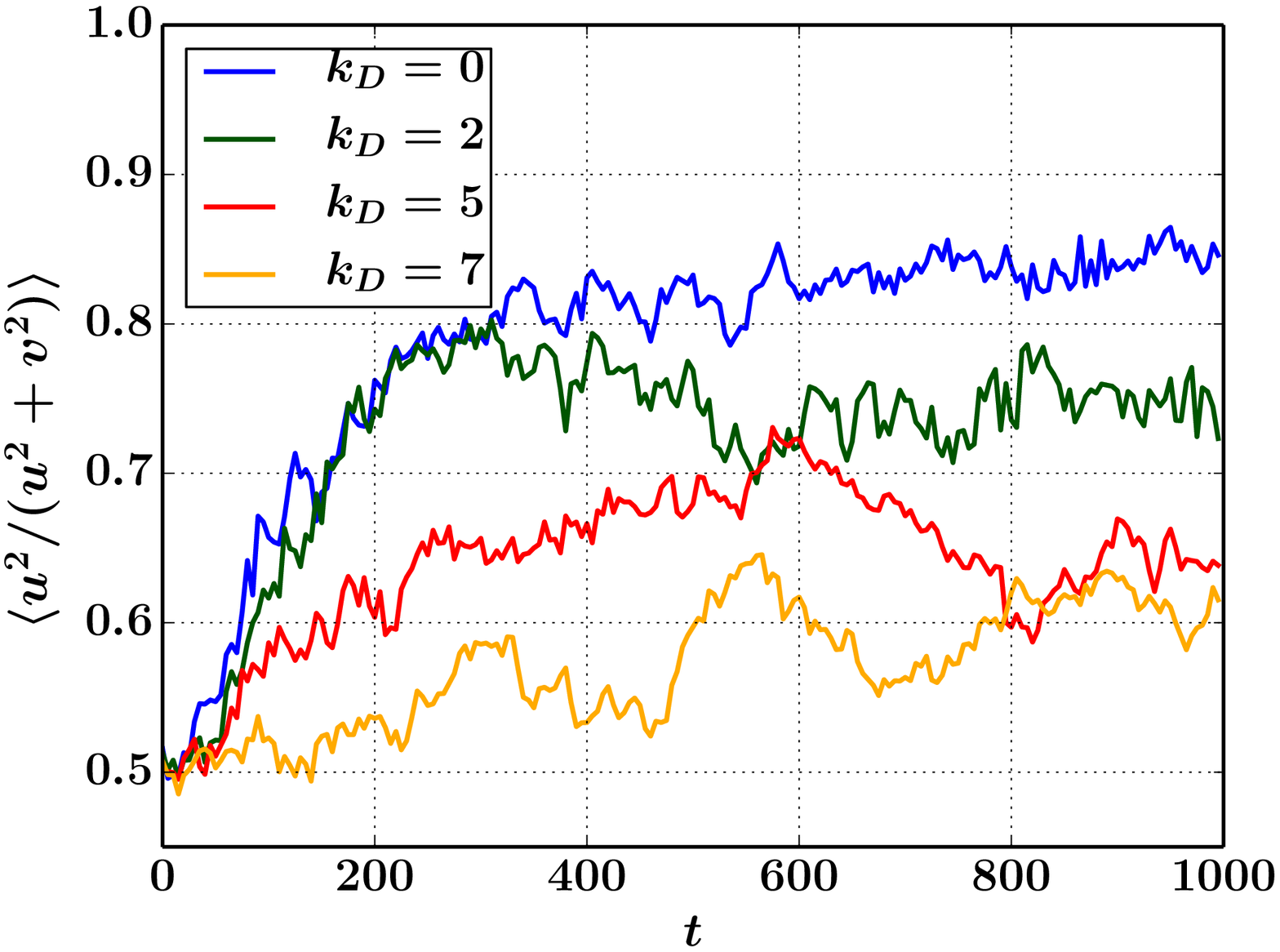}
\includegraphics[width=7cm,height=6cm]{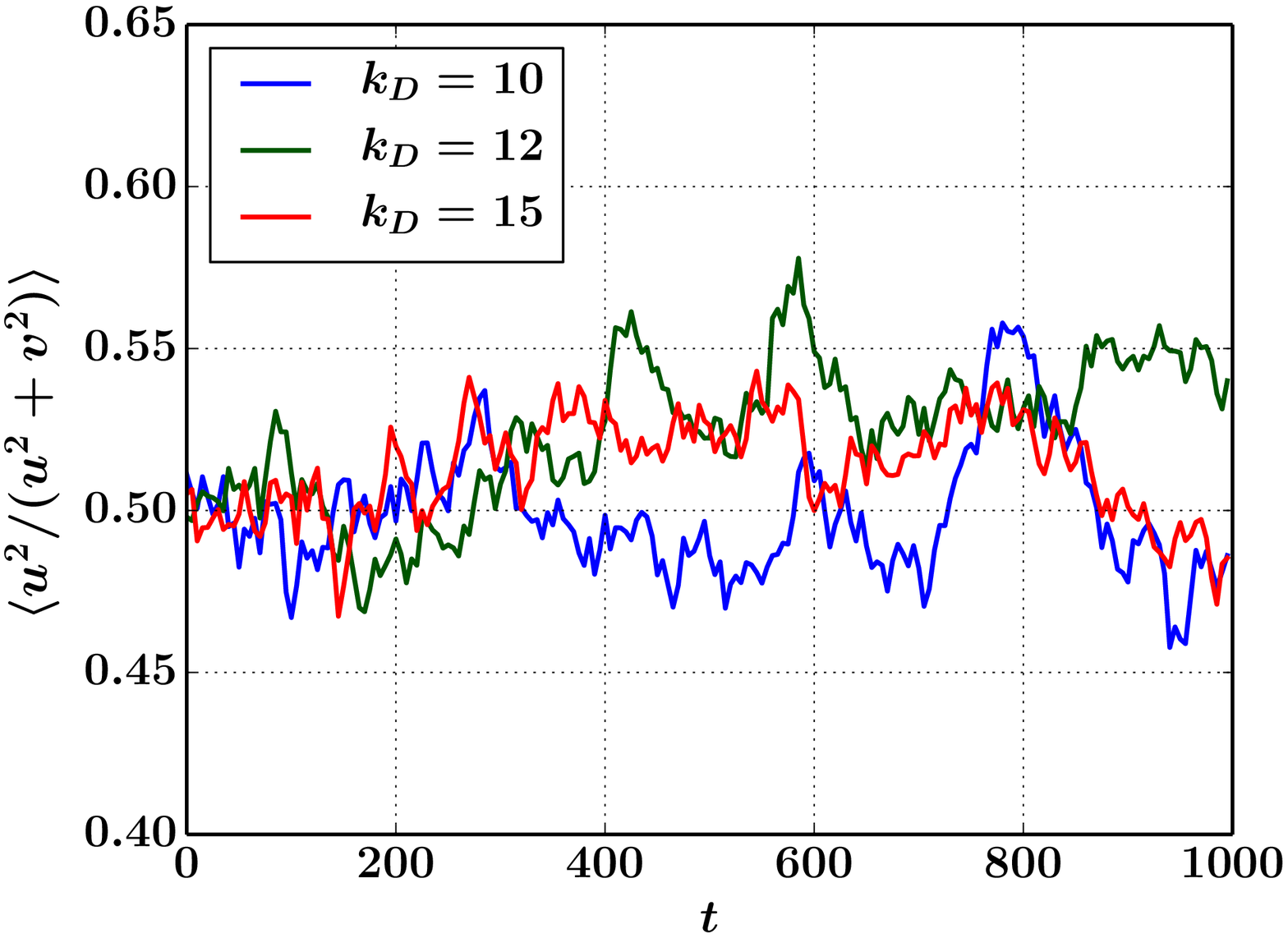}
\caption{\label{anisotropy} Anisotropy of various runs ($k_D = 1/L_D$). Panel (a) $L_D=\infty, 1/2, 1/5$ and $1/7$. Panel 
(b) $L_D=1/10, 1/12$ and $1/15$.}
\end{figure}

\begin{figure}
\centering
\includegraphics[width=7cm,height=6cm]{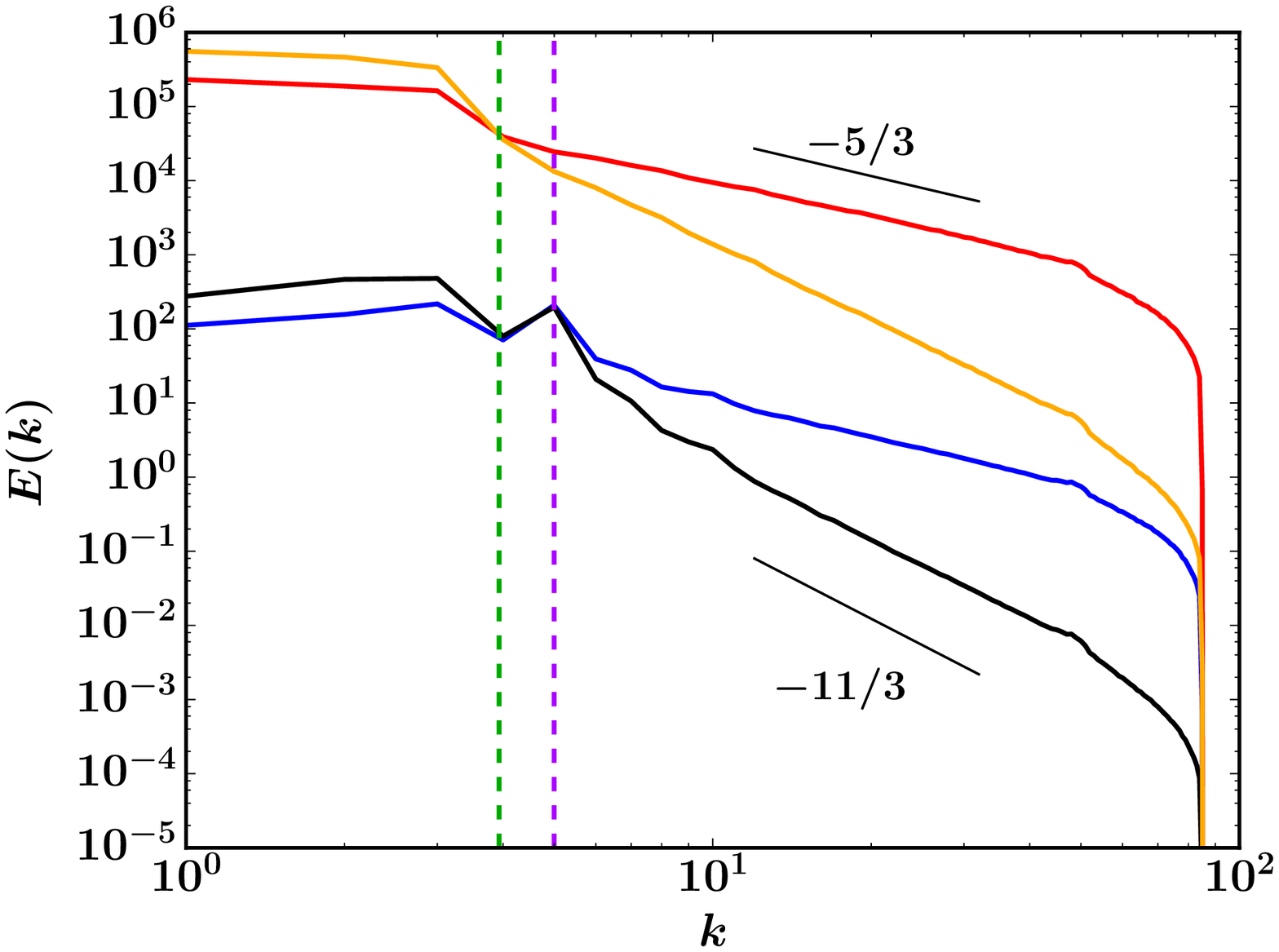}
\includegraphics[width=7cm,height=6cm]{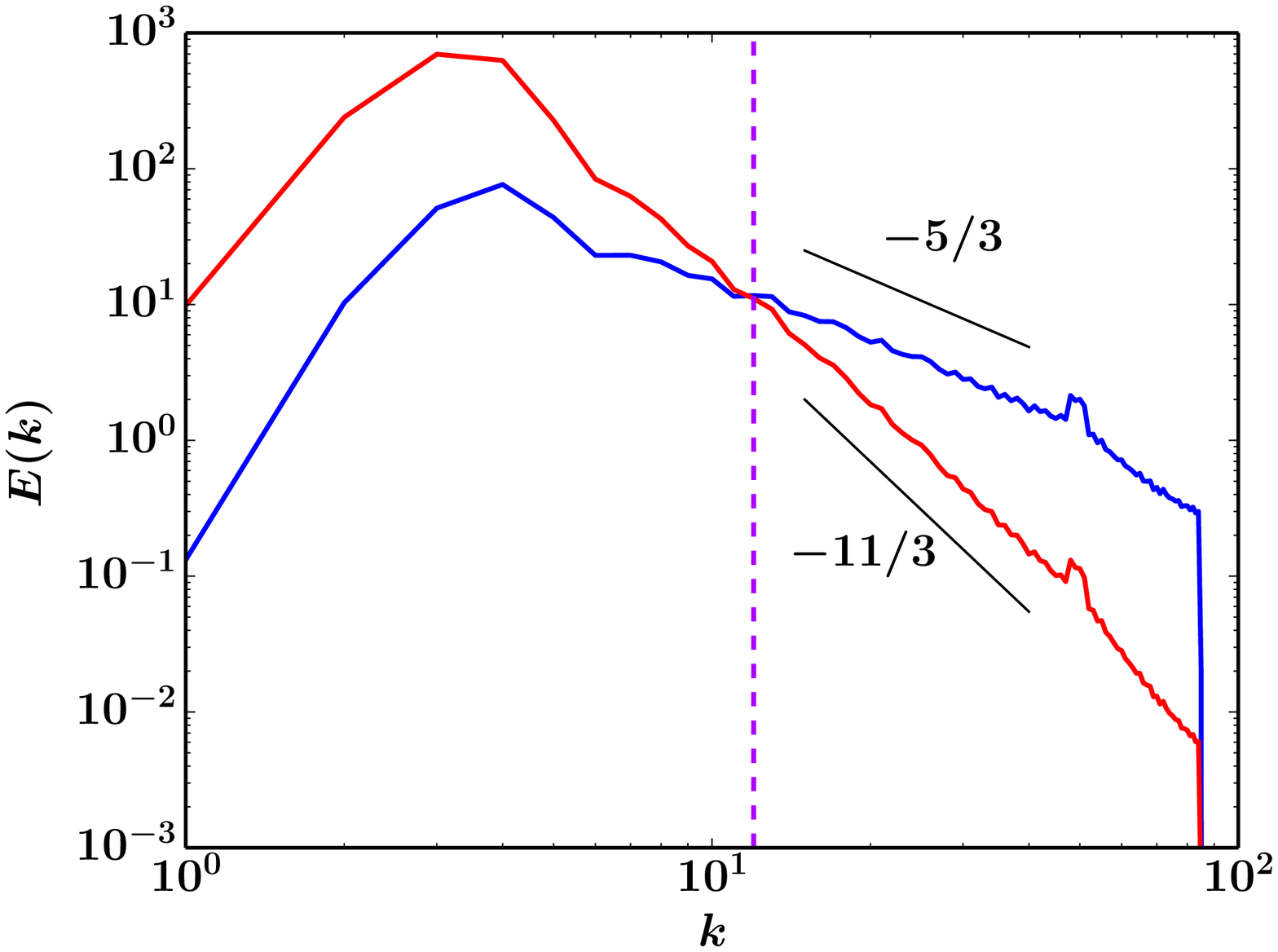}
\caption{\label{ekspectra} KE and PE spectra. Violet and green dashed lines are the deformation scale and transition scale ($k_T=1/L_T$). 
Panel (a) shows the spectra with respect to $k_x$ (red and orange, shifted for clarity) and $k_y$ (blue and black) for $k_D=5$ (i.e., with jets). Panel (b) 
shows a no jet case ($k_D=12$), as the spectra are isotropic in this case they are shown with respect to $k$. Note that the spectra cross at $k_D$ and follow
power-laws close to -5/3 (KE) and -11/3 (PE). }
\end{figure}

Inspecting the anisotropy of the various runs (Figure \ref{anisotropy}), i.e., the ratio $\langle u^2/(u^2+v^2) \rangle$ (where $\langle \cdot \rangle$ denotes a spatial average),
we observe clear zonal flow domination or jet formation for $L_D=\infty,1/2$ and 1/5, with an anisotropy of 0.85, 0.75 and 0.65 respectively.
As expected, the jets are asymmetric in the east-west direction and are associated with a ramp like
structure in the vorticity field (not shown).
In each of these cases the Rhines scale ($L_R$) is approximately 1/5, and we observe 5-6 jets in the system
at equilibrium. Also, once the jets form they are persistent in the sense their number remains the same and their zonal mean profile is
qualitatively similar.
The formation of jets is consistent with criterion (\ref{3}) derived by Okuno and Masuda \cite{Okuno}. In addition,
based on the energy input to the system, we note that whenever jets form, the critical $\beta$ criterion of Smith \cite{Smith-2004}, namely
$\beta (\epsilon k_d^5)^{-1/3} \ge 3.9$ (where $\epsilon$ is the energy input rate) is also satisfied. It is known that the spectra in jet forming cases 
are anisotropic (see for example, Sukoriansky et al. \cite{SDG-2007}), in particular, with respect to $k_y$
(illustrative case of $L_D=1/5$ shown in the first panel of Figure \ref{ekspectra}), the slope steepens as one progresses to 
lower wavenumbers and we observe a ``piling up" of energy
at scales that are slightly greater than the Rhines scale. In fact, the
transition scale ($L_T$) --- also called the modified Rhines scale \citep{Okuno} --- turns out to be a reasonable measure of this length scale.
On the other hand, the scaling with respect to $k_x$ is smoother (and proceeds to largest scales available where it is arrested by the drag imposed on the simulation) with the kinetic energy (KE) and potential energy (PE) following a -5/3 and -11/3
power-law, respectively. 

The case $L_D=1/7$ is somewhat borderline
with an anisotropy of 0.6 and for $L_D=1/10,1/12$ and 1/15 there are no jets, rather the fields consist of small-scale isotropic vortices and the anisotropy
fluctuates about 0.5. In conformity with Okuno and Masuda \cite{Okuno} and Smith \cite{Smith-2004}, these cases do not satisfy their jet formation criteria. In fact, with progressively
smaller deformation scales, the emergence of isotropic vortices (which do not merge, but retain their small scales) is consistent with the
behavior of the $L_D \rightarrow 0$ limit of the equivalent barotropic
model \citep{Larichev-Mc,ISW-2002}.

\begin{figure}
\centering
\includegraphics[width=7cm,height=6cm]{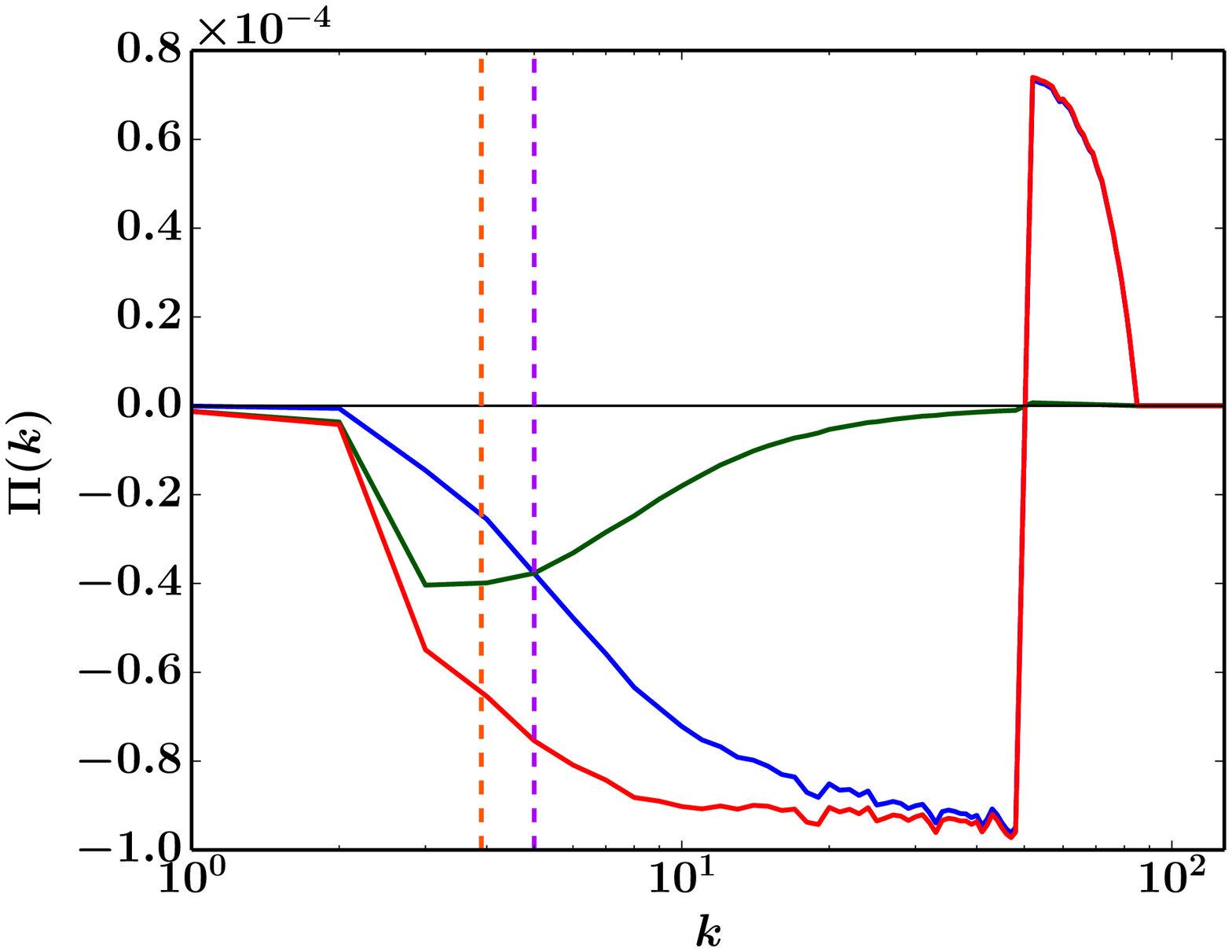}
\includegraphics[width=7cm,height=6cm]{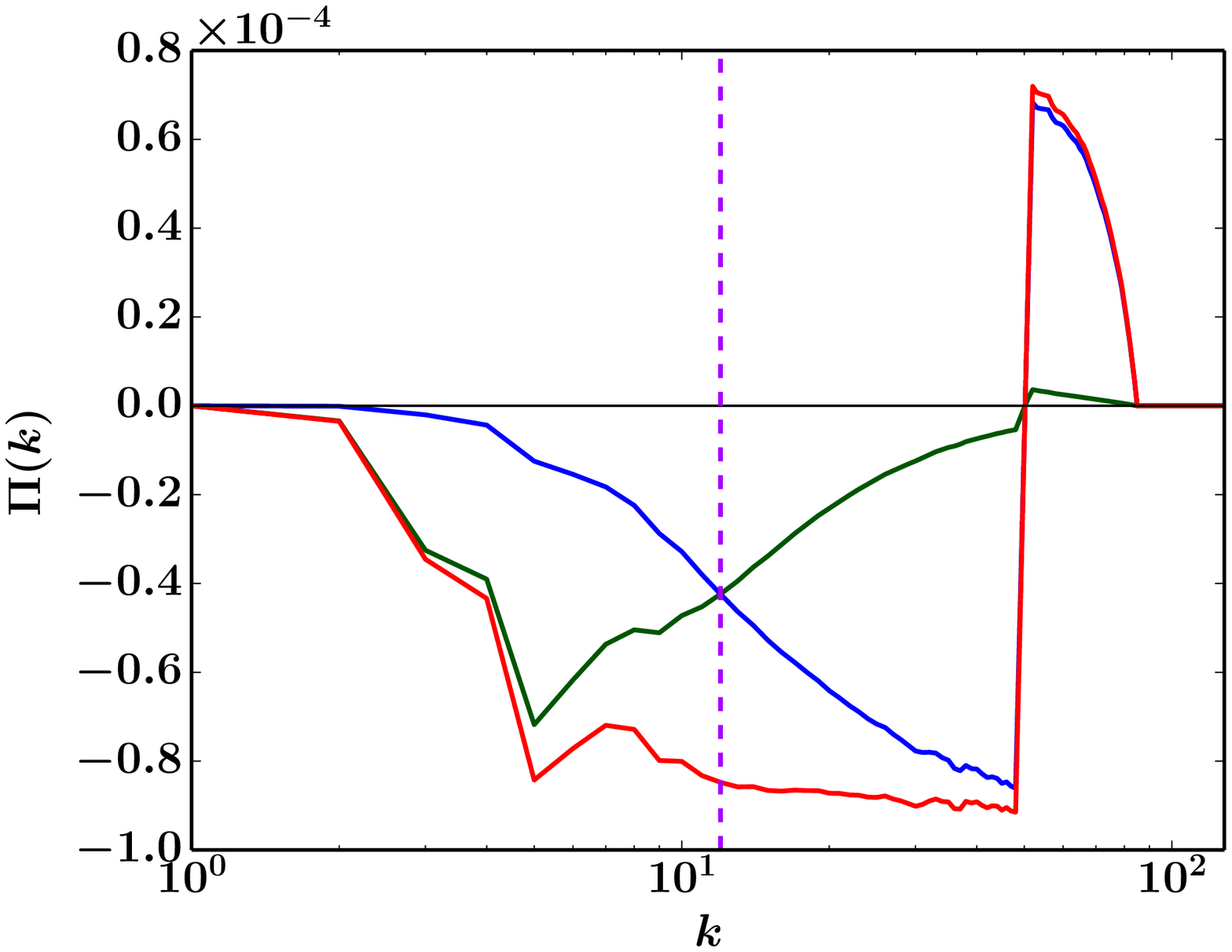}
\caption{\label{flux} Flux (averaged over 100 time units) in wavenumber domain. Blue, green and red solid lines are respectively the KE, PE and total energy flux. Violet and orange dashed lines are the deformation scale and transition scale. Panel (a) shows $k_D=5$ (jets) and (b) shows $k_D=12$ (no jets).}
\end{figure}

Irrespective of whether jets form or not, there is a systematic inverse transfer of kinetic and potential energy. As
noted earlier \citep{Scott2}, the spectra cross at $L_D$
(illustrative examples with and without jets are shown in the two panels of Figure \ref{ekspectra}) with the
KE and PE
dominating for scales smaller and larger
than the deformation scale, respectively. Note that the spectra in the non jet case are isotropic, and have been shown with respect to $k$.
The flux of KE falls off by the deformation scale (Figure \ref{flux}), while the PE dominates the inverse flux at scales larger
than $L_D$. In fact, the flux of total energy is relatively steady over a fairly large range of scales, 
and its spectrum
follows the PE for $k<1/L_D$ and KE for $k>1/L_D$, thus exhibiting a natural transition from a steep (-11/3) to shallow (-5/3) spectrum at $L_D$
(this is especially clear in the non jet
forming cases, as seen in the illustrative plot in the second panel of Figure \ref{ekspectra}).

\subsection{Analysis in frequency domain}

\begin{figure}
\centering
\includegraphics[width=7cm,height=6cm]{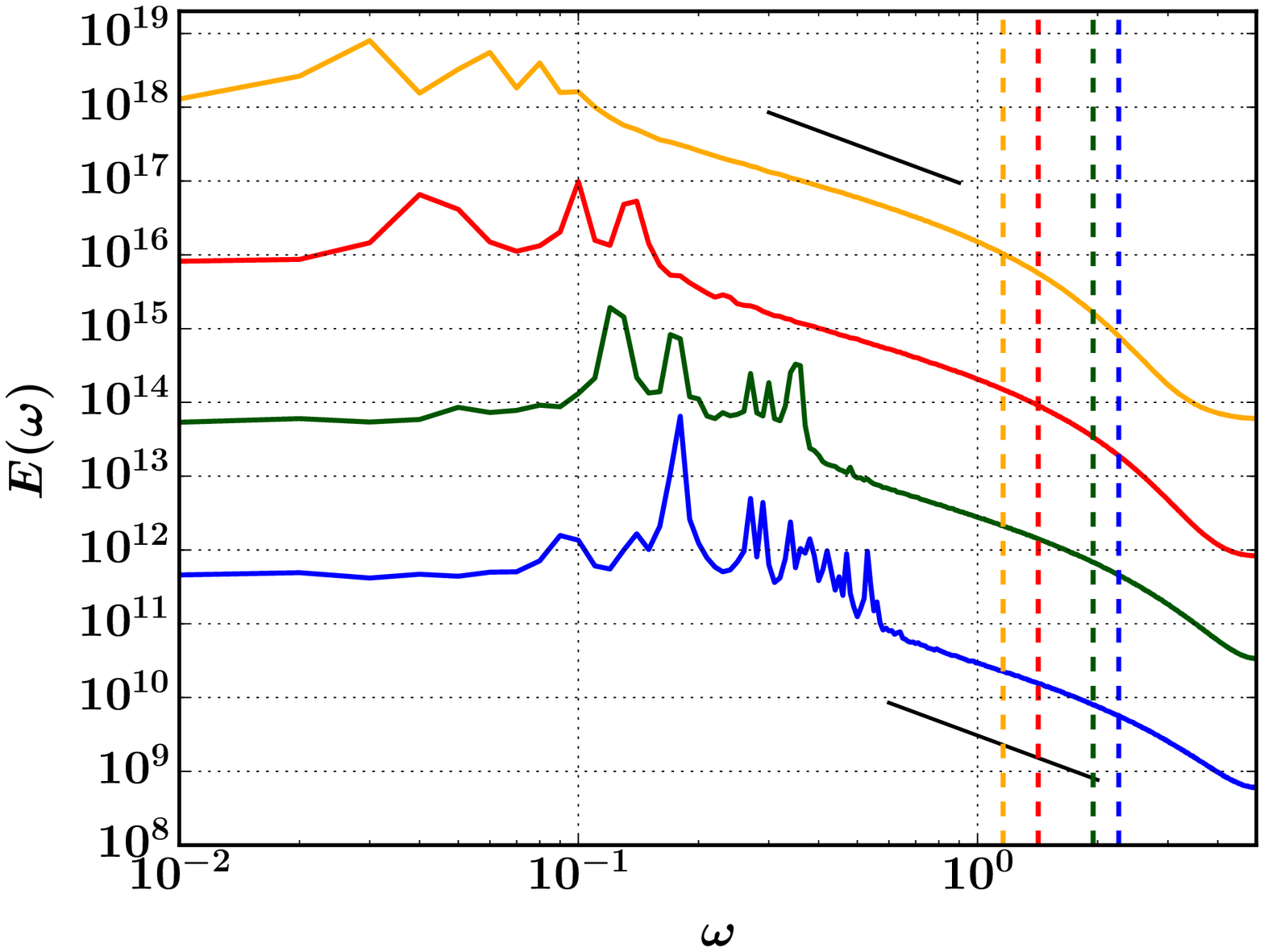}
\includegraphics[width=7cm,height=6cm]{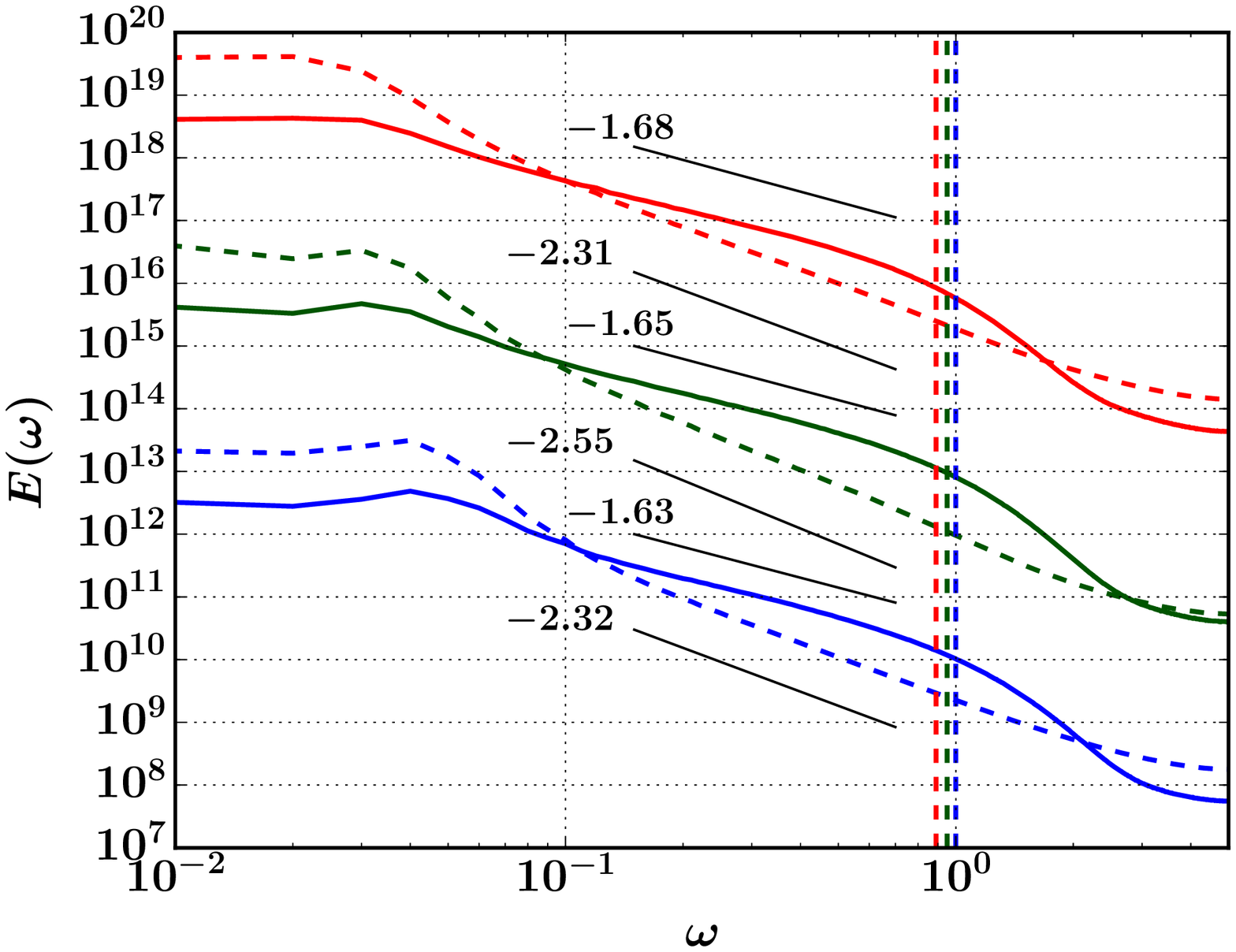}
\caption{\label{Envsfreq} Energy spectra in frequency domain. Panel (a) shows the KE in cases where jets form 
(blue, green, red and orange curves represent $k_D=0,2,5$ and $7$). Black line denotes a slope of 2. Panel
(b) shows the KE (solid lines) and PE (dashed lines) for non-jet forming cases 
(blue, green and red curves represent $k_D=10,12$ and $15$). The best fit slope is shown in each case. 
Vertical dashed lines denote the frequency of forcing for the correspondingly colored spectra. 
Note that the spectra are averaged for four 100 time period samples ($t=800-900, 900-1000, 1000-1100, 1100-1200$) and have been shifted for clarity.}
\end{figure}

Following Tennekes and Lumley \cite{Ten} and Landau and Lifshitz \cite{Landau}, for a wavenumber scaling of -5/3, the scaling of a cascade in the 
frequency domain is expected to follow a -2 power law. This is in contrast to associating a ``frequency" with 
turbulence ($U k$; equivalent to invoking the Taylor hypothesis), 
that would result in the same functional
form in wavenumber and frequency domains (see for example the discussion of wavenumber and frequency spectra in
Ferrari and Wunsch \cite{FW-2010}).
As seen in Figure \ref{Envsfreq}, there is some
evidence of
$\omega^{-2}$ scaling in the KE, though it only lasts for a very limted range. Further, for larger deformation scales (the jet forming cases in the first panel of Figure \ref{Envsfreq}), i.e., $L_D=\infty,
1/2$ and $1/5$, there are systematic departures from a power-law for progressively smaller frequencies. Note that, for these deformation scales, the PE contains a 
much smaller energy at higher frequencies and its spectrum shows no clear power-law. 
These deviations in KE are in the form of isolated
peaks, and their decreasing frequency (with increasing $L_D$) is consistent with the Rossby dispersion relation. Not only do the peaks 
shift to lower frequencies with a decreasing deformation scale, they become muted, and in fact, the power-law scaling of KE in the non-jet forming 
cases is much clearer and persists to the lowest frequencies sampled (second panel of Figure \ref{Envsfreq}). Also, the best-fit 
slopes for KE 
in the non-jet forming cases are close to -5/3  (this is in accord with simulations of the pure 2D
system (not shown), i.e., $\beta=0, L_D=\infty$, that yield an extensive power-law that scales almost exactly with a -5/3 slope in frequency and 
wavenumber space). 
Interestingly, the PE in the 
non-jet forming simulations falls off quite rapidly at low frequencies and levels off to a relatively steep slope 
(steeper than -2), as compared to the KE, for higher frequencies (dashed lines in the second panel of Figure \ref{Envsfreq}).

Thus, we have an interesting set of observations wherein the scaling of the KE with respect to frequency for jet forming 
cases (at least at large frequencies) shows some evidence of a steeper than -5/3 (possibly consistent with a -2 slope) power-law though the 
scaling is delicate as the range of its validity is 
very small, while the isotropic non-jet forming cases appear to 
follow Taylor's hypothesis and exhibit a fairly clear $\approx$ -5/3 scaling in KE but show a steeper
slope in the PE (note that even though the PE slope is steeper than the KE, the two do not differ by -2 as was seen in the wavenumber spectra). Indeed, much like wavenumber space, the total energy in the non-jet forming cases again exhibits a transition from a steep (PE dominated)
to shallow (KE dominated) form with increasing frequency. The transition point of the KE and PE spectra is in fair agreement with a straightforward 
translation of the deformation
scale to the frequency domain by using the the observed $U$ as a conversion factor.

\begin{figure}
\centering
\includegraphics[width=7cm,height=6cm]{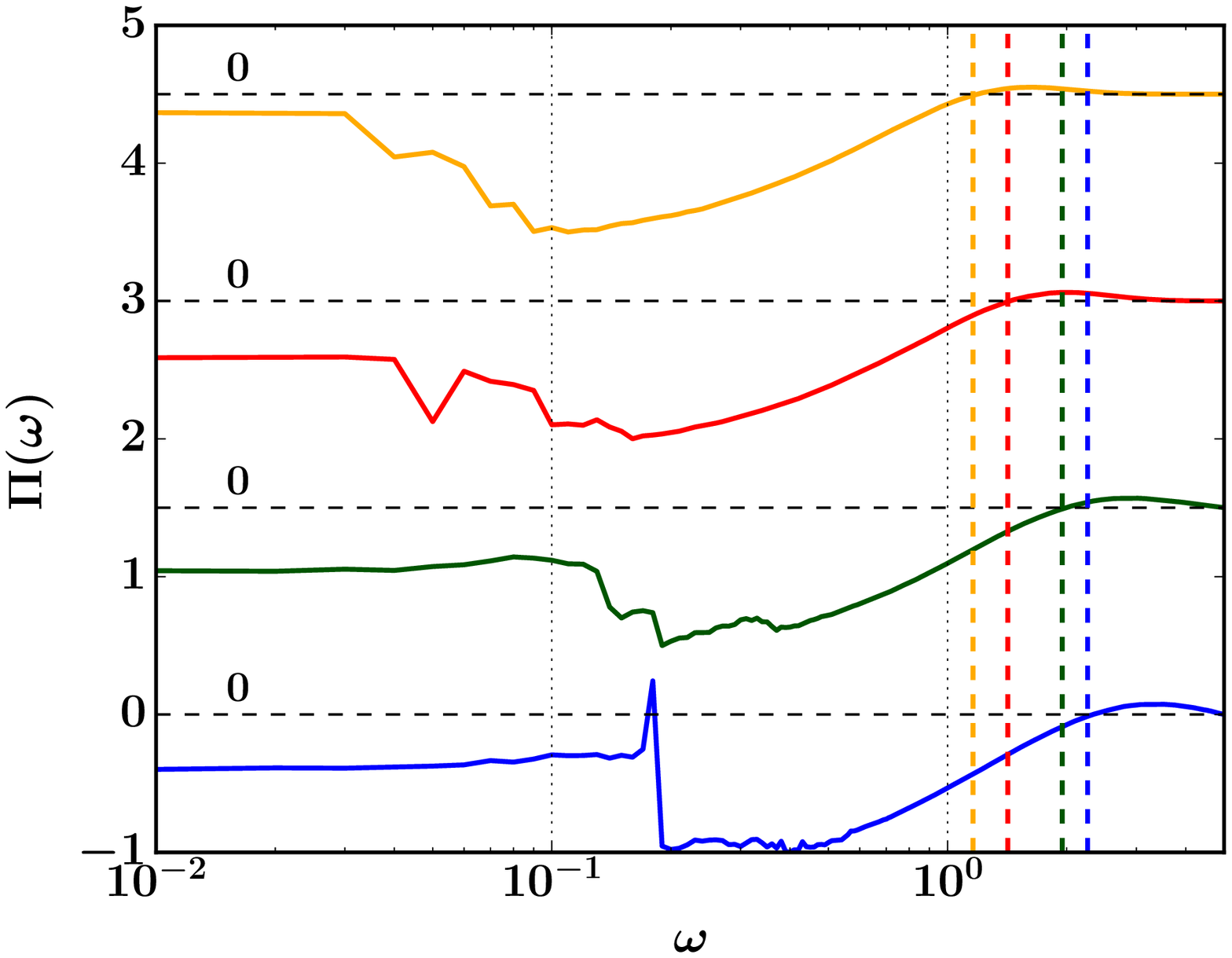}
\includegraphics[width=7cm,height=6cm]{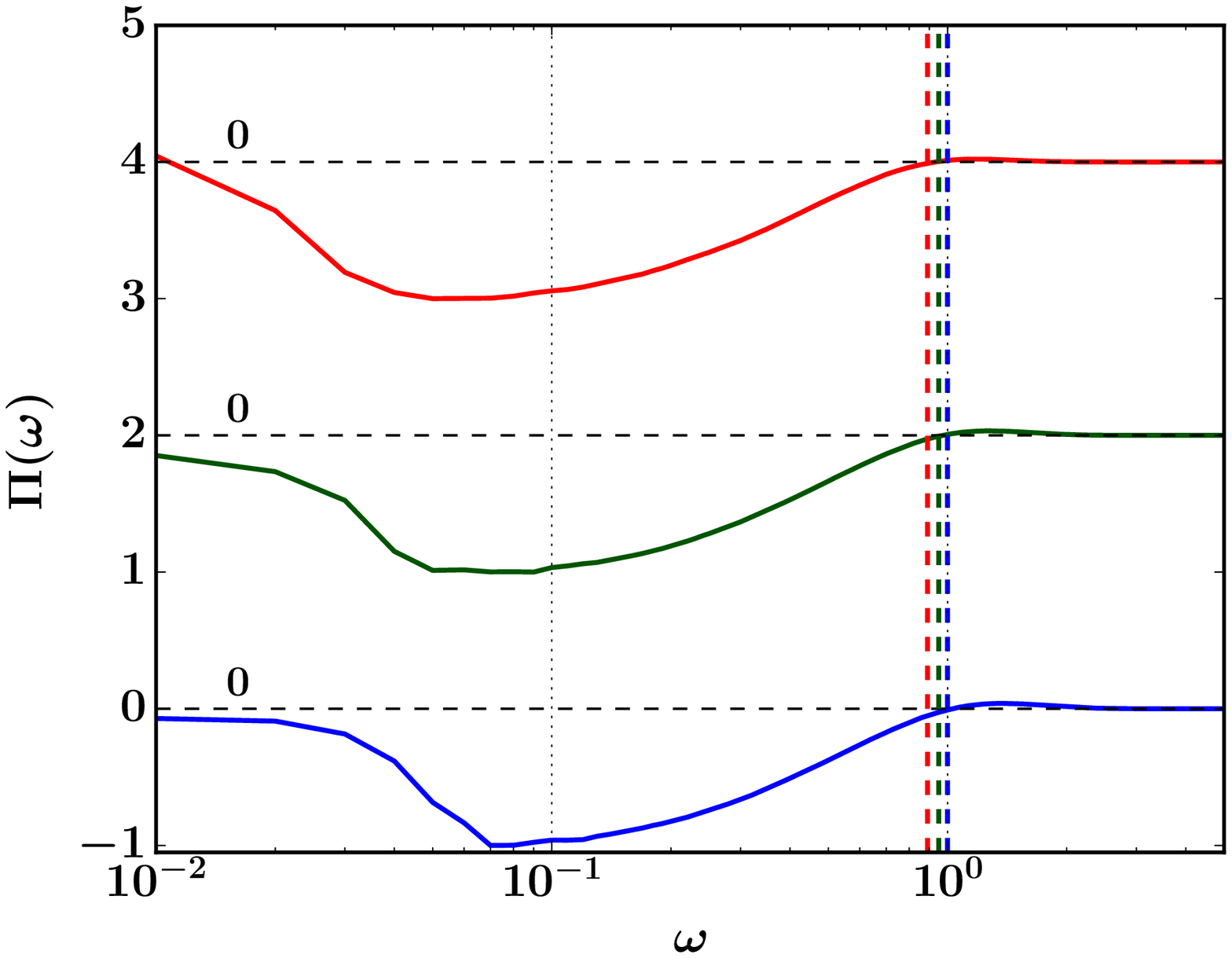}
\caption{\label{freqflux} Normalized KE flux in frequency domain. Panel (a) and (b) follow the notation of Figure \ref{Envsfreq}. 
The fluxes have been shifted for clarity and
the mean of four 100 time period samples ($t=800-900, 900-1000, 1000-1100, 1100-1200$) is plotted.}
\end{figure}

In all cases, there is a flux of KE from larger to smaller frequencies, this is seen in Figure \ref{freqflux}. Note that the inverse flux in the
frequency domain
is broader than in wavenumber space \citep{Arbic-2012}. Interestingly, comparing Figure \ref{Envsfreq} and \ref{freqflux}, we
see that the power-law scaling of KE does not correspond to an ``inertial range" in the flux, in fact, the scaling only lasts as long as the flux keeps decreasing.
In both, Figure \ref{Envsfreq} and Figure \ref{freqflux}, the vertical dotted lines indicate the frequency of the forcing ($U k_f; k_f=50$), and as is evident,
the inverse transfer begins at the forcing frequency itself.

\subsection{Frequency-zonal wavenumber diagrams}

\begin{figure}
\centering
\includegraphics[width=7cm,height=6cm]{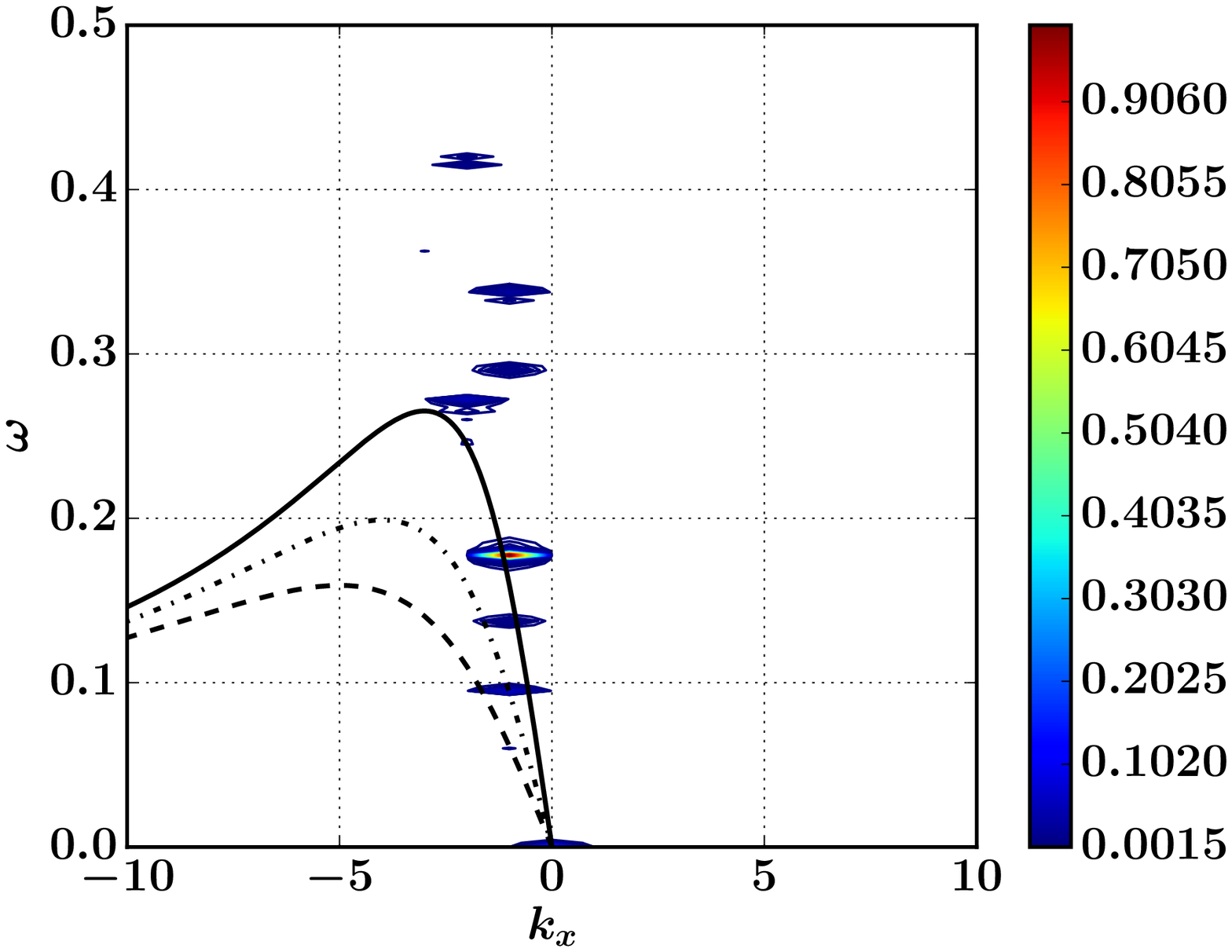}
\includegraphics[width=7cm,height=6cm]{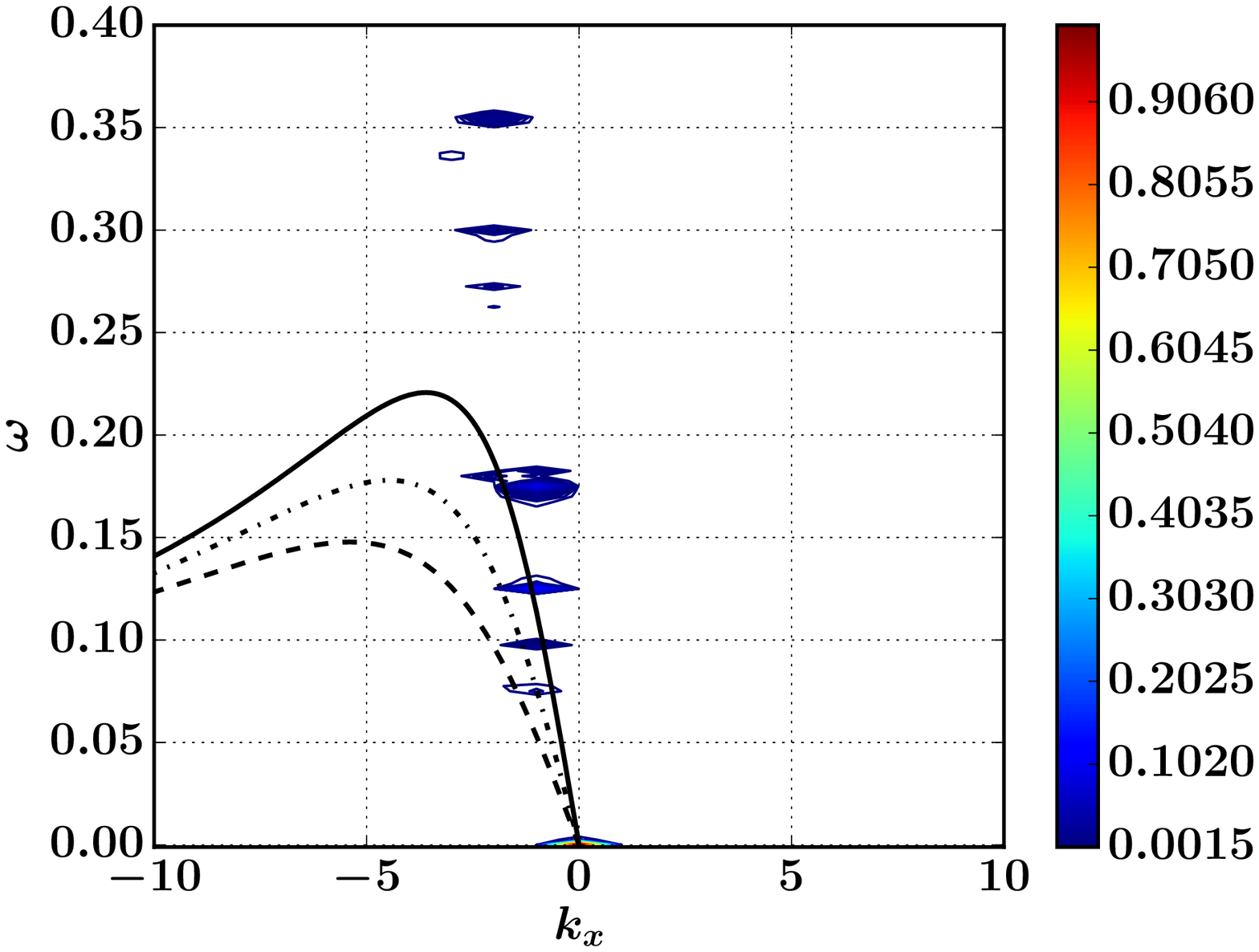}
\includegraphics[width=7cm,height=6cm]{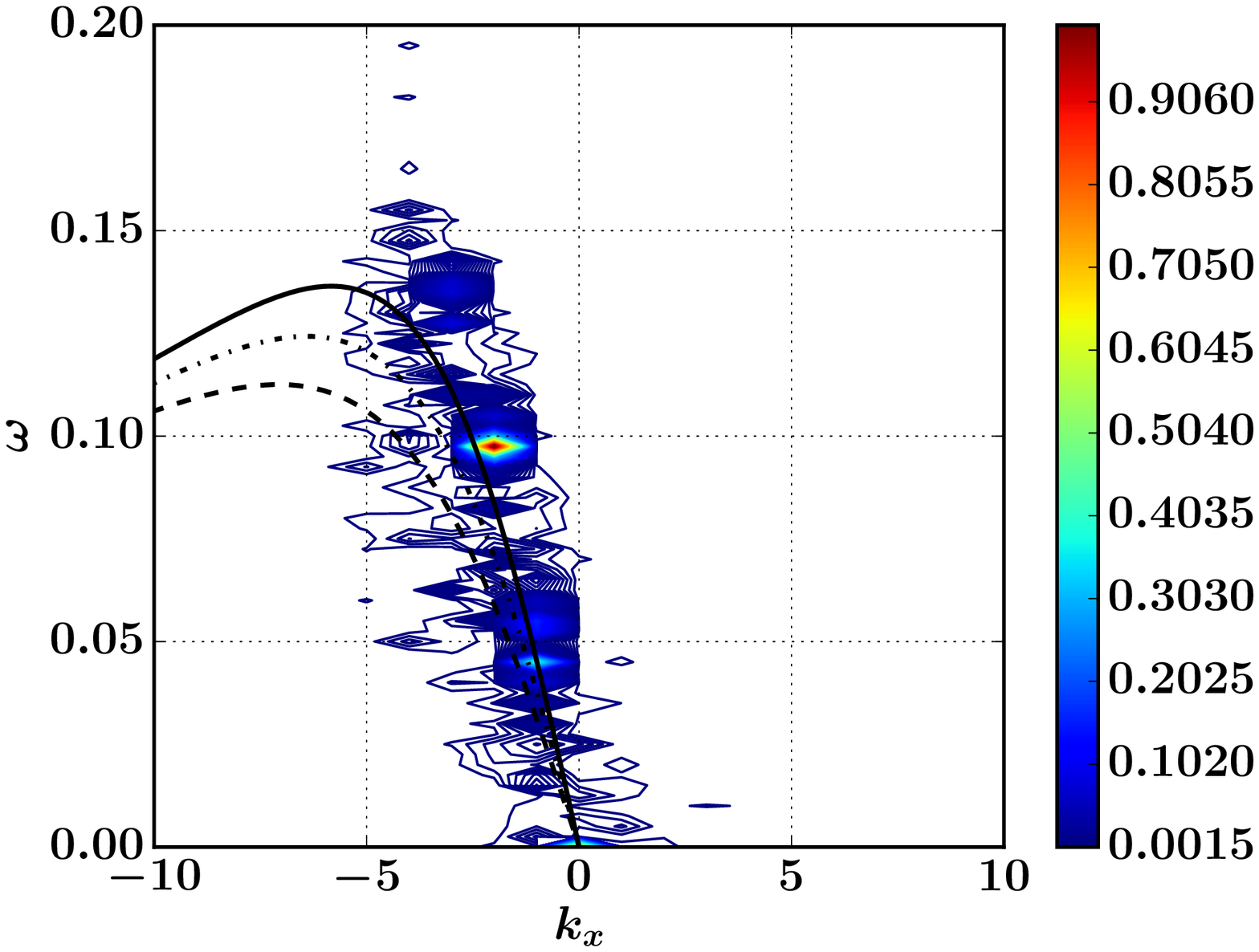}
\includegraphics[width=7cm,height=6cm]{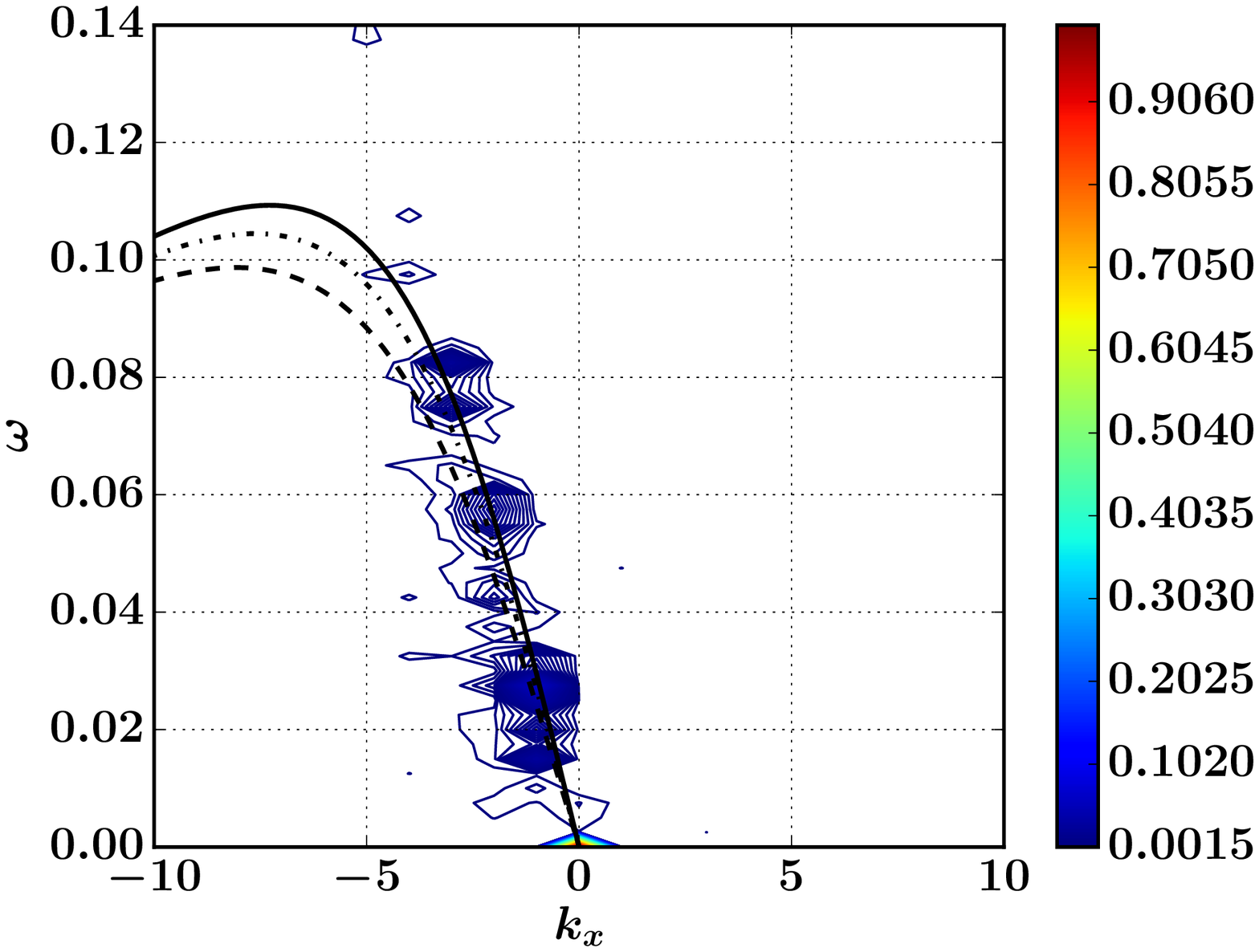}
\caption{\label{fkx_jets}  Frequency-zonal wavenumber plots constructed using 4000 sequential snapshots of the zonal velocity for $k_D=0,2,5$ and $7$ in the top left, top right, bottom left and bottom right panels, respectively. Solid, dash-dot and dashed curves are the Rossby dispersion relations for $k_y=k_T-1, k_T$ and $k_T+1$ ($k_T = 4$ for $k_D = 0,2,5$ and $k_T=3$ for $k_D=7$).}
\end{figure}

To ascertain the character of the disturbances represented by peaks in Figure \ref{Envsfreq}, we examine the distribution of energy in frequency-wavenumber diagrams.
As noted, when jets form, $L_T$ serves as a useful indicator of the meridional scale at which energy tends to pile up (first panel of Figure \ref{ekspectra}). Thus, fixing $1/L_T$ as a meridional
wavenumber, we examine the energy distribution in frequency-zonal wavenumber space. 
As described earlier, these
diagrams are constructed by computing the spectra of longitude-time plots of the zonal velocity at a given latitude for multiple snapshots of the
flow field.
These diagrams for $L_D=\infty, 1/2, 1/5$ and 1/7
are shown in the four panels of Figure \ref{fkx_jets}. In each case,
we observe a peak at $k_x \approx 0$ and $\omega \approx 0$; this is the steady zonal jet. In addition, there are peaks for small (negative) $k_x$ and
low (but, positive non zero) frequencies that indicate the presence of westward moving disturbances. Further, the peaks observed are consistent 
with low zonal wavenumber theoretical Rossby dispersion curves corresponding to (\ref{1}) with $k_y = 1/L_T$ and $k_y=1/L_T \pm 1$ (shown as black 
curves in Figure \ref{fkx_jets}).
Note that there appears to be a slight increase in $k_x$, or decrease in scale, of the modulating waves with
progressively smaller deformation scales; for example, with $L_D=\infty$ the spectral peaks are restricted to $k_x < 2$, while for $L_D=1/7$, 
we see additional 
peaks out to $k_x \approx 5$. Also, the separation between the jet ($k_x, \omega \approx 0$) and the other peaks tends to
diminish with a decreasing deformation scale. Indeed, by $L_D=1/7$ (the borderline jet case), there is almost a continuous distribution of energy from
zero to low frequencies. Quite clearly, the energy in these frequency-zonal wavenumber diagrams is strongly localized; in fact, one could argue that a linear
dispersion relation would fit the peaks as well as Rossby dispersion curves. Therefore, we postpone the identification of these disturbances to the next section 
when we have also analyzed 
the frequency-meridional wavenumber diagrams.

\begin{figure}
\centering
\includegraphics[width=7cm,height=6cm]{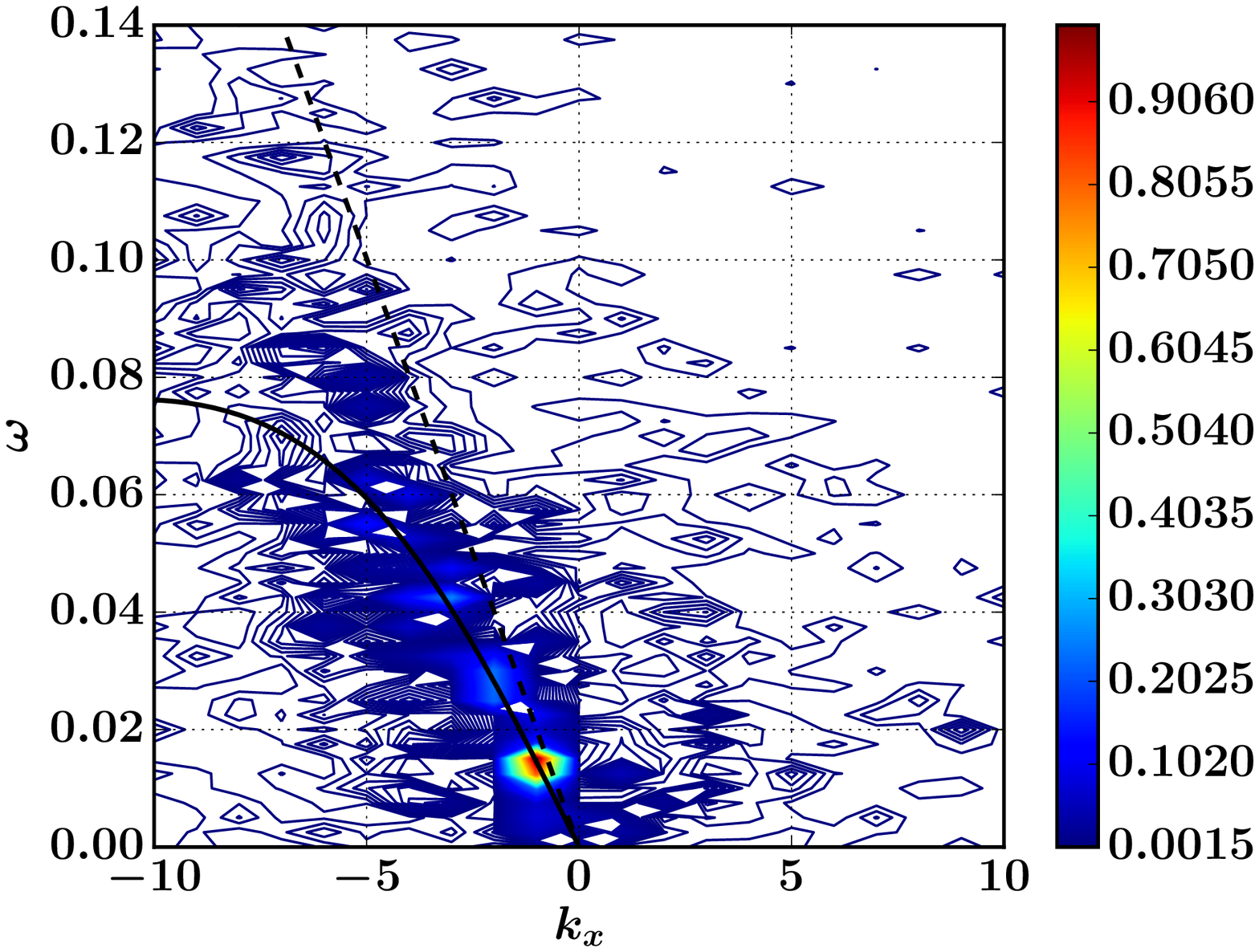}
\includegraphics[width=7cm,height=6cm]{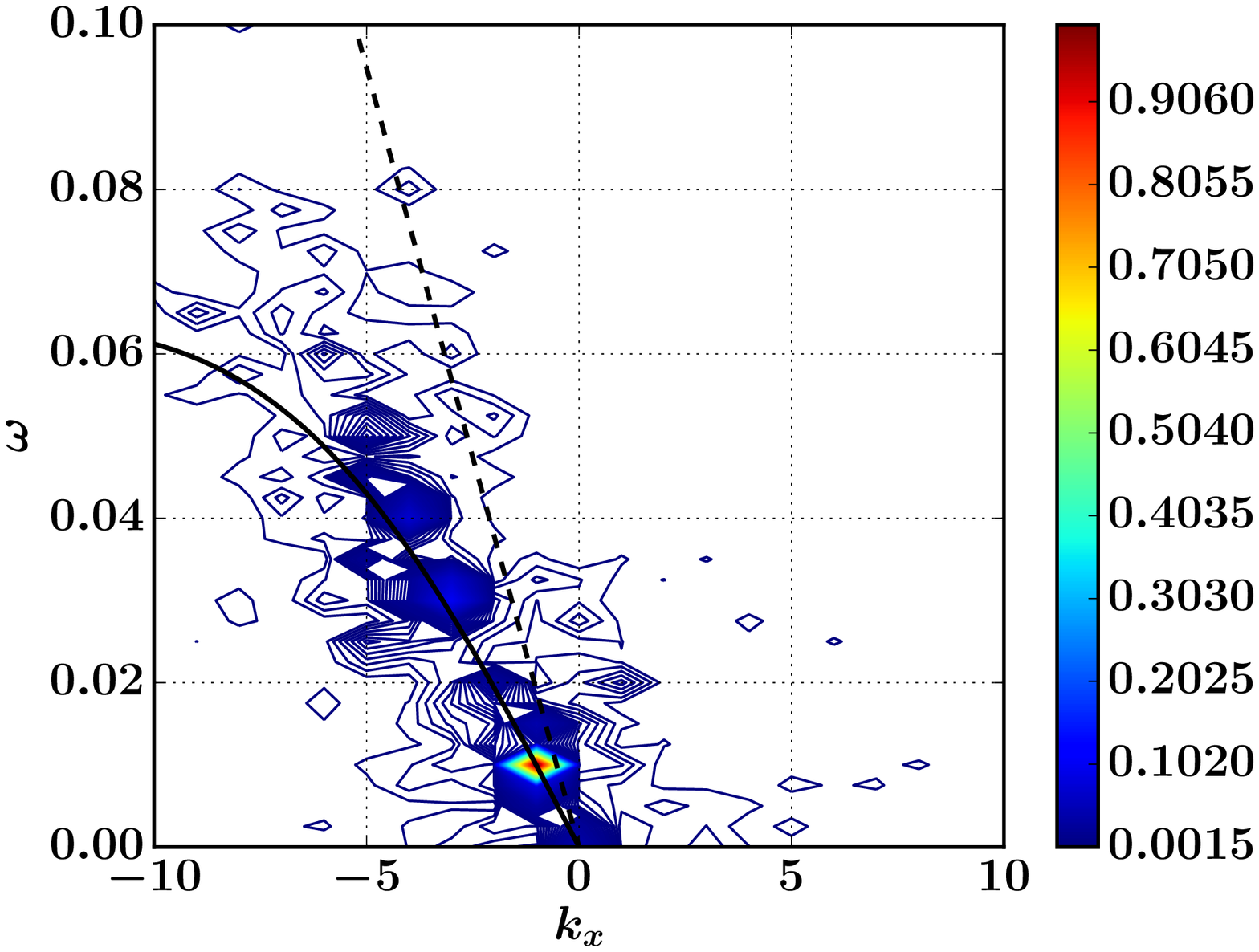}
\includegraphics[width=7cm,height=6cm]{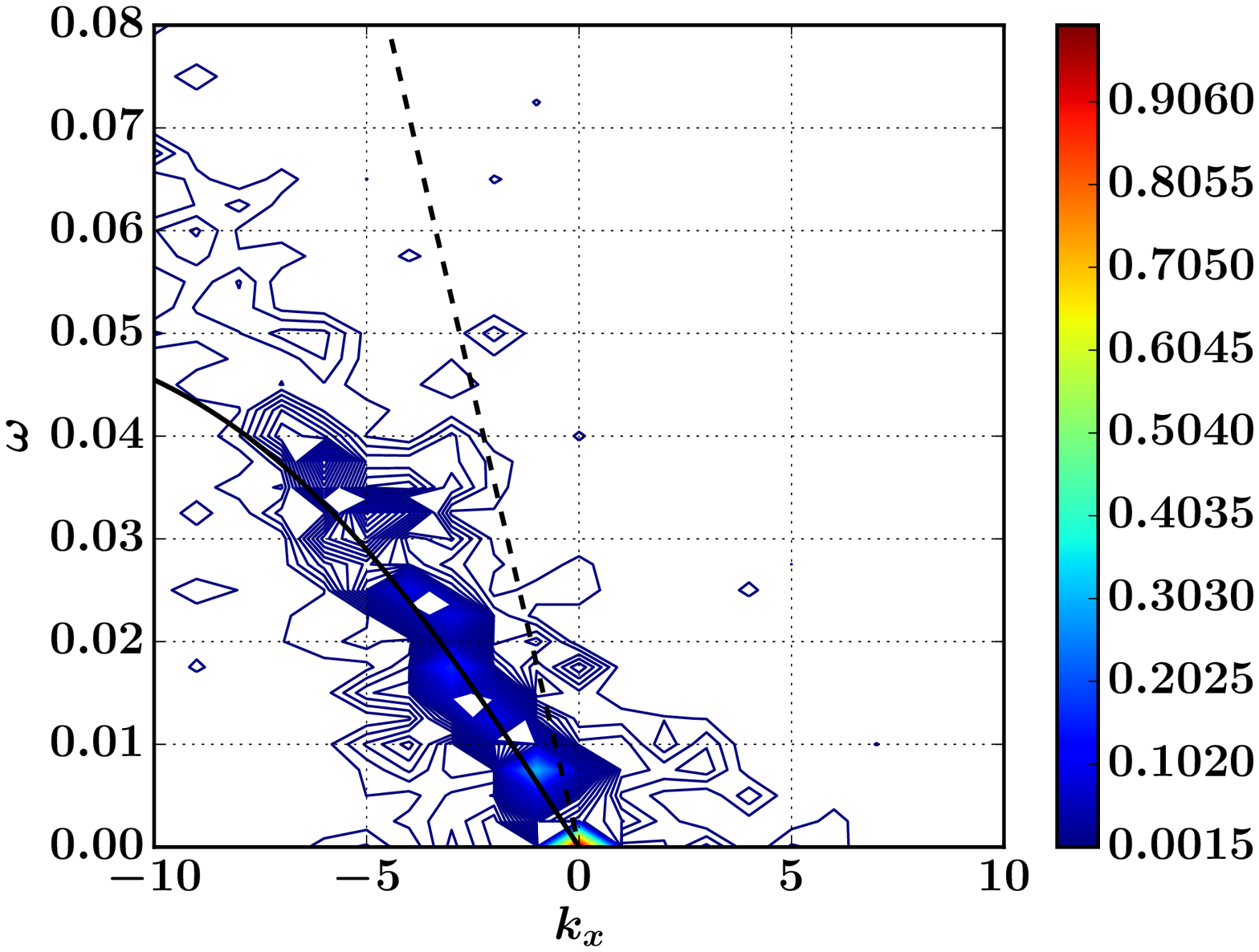}
\caption{\label{fkx_nojets} Frequency-zonal wavenumber plots constructed using 4000 sequential snapshots of the zonal velocity for $k_D=10,12$ and $15$ in
the top left, top right and bottom panels, respectively. Solid and dashed curves are the Rossby dispersion relation and
turbulence frequencies ($k_y = 3,4$ and $5$ for $k_D=10,12$ and $15$). Also shown are dispersion lines for ``turbulence", i.e., $Uk_x$.}
\end{figure}

Moving to the cases when jets do not form, i.e., $L_D=1/10, 1/12$ and 1/15, the energy distribution in frequency-zonal wavenumber space is shown in
Figure \ref{fkx_nojets}. Here too, most of the power is in westward propagating systems and there is a possibility that the observed peaks could be associated with Rossby 
dispersion curves.
As there is no meridional length scale to guide us, we have plotted the
dispersion curves with $k_y$ that best fits the peaks in the spectra.
In comparison with the jet forming simulations, the energy is quite broadly distributed and not closely confined to the
dispersion curves.

\subsection{Frequency-meridional wavenumber diagrams}

\begin{figure}
\centering
\includegraphics[width=7cm,height=6cm]{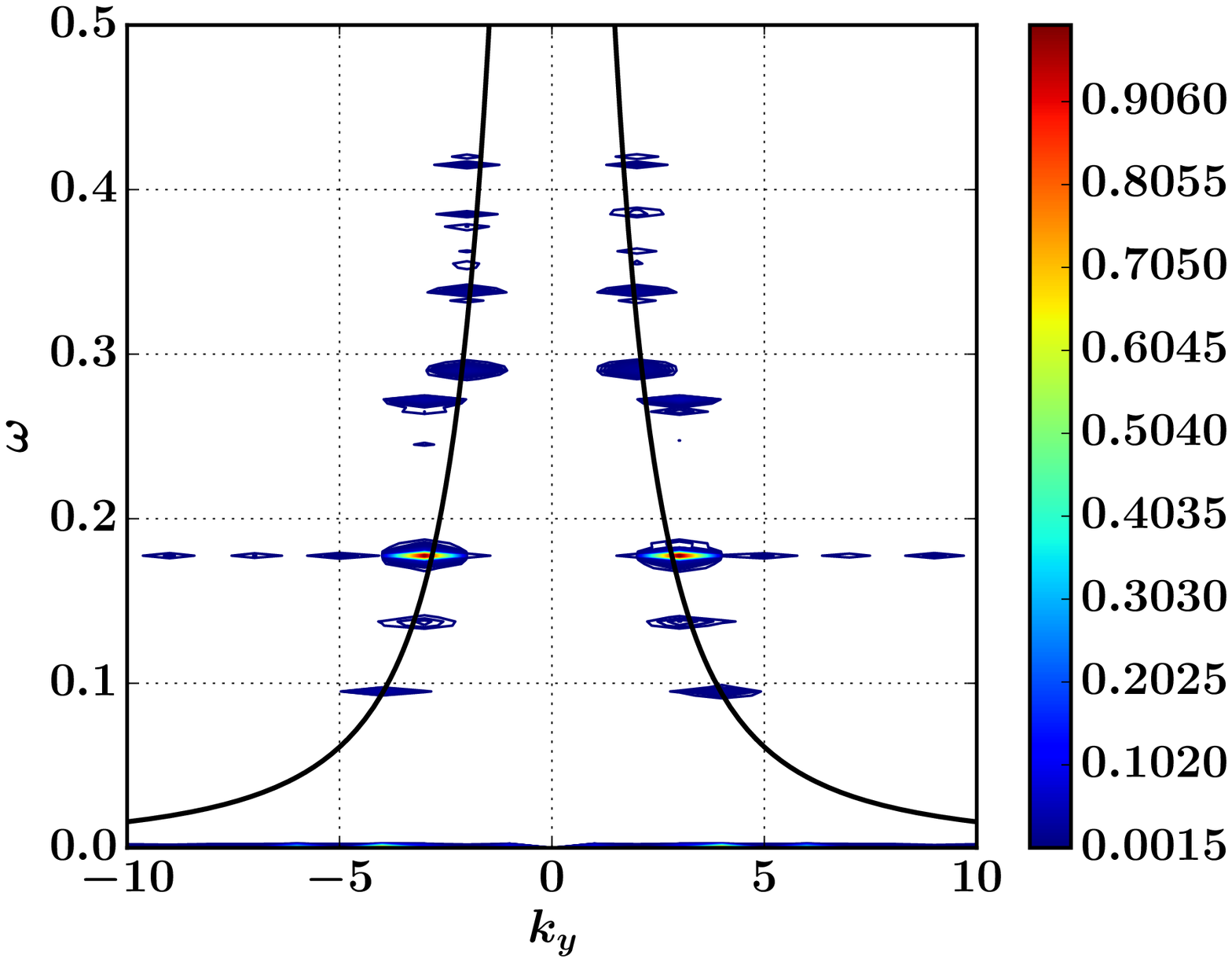}
\includegraphics[width=7cm,height=6cm]{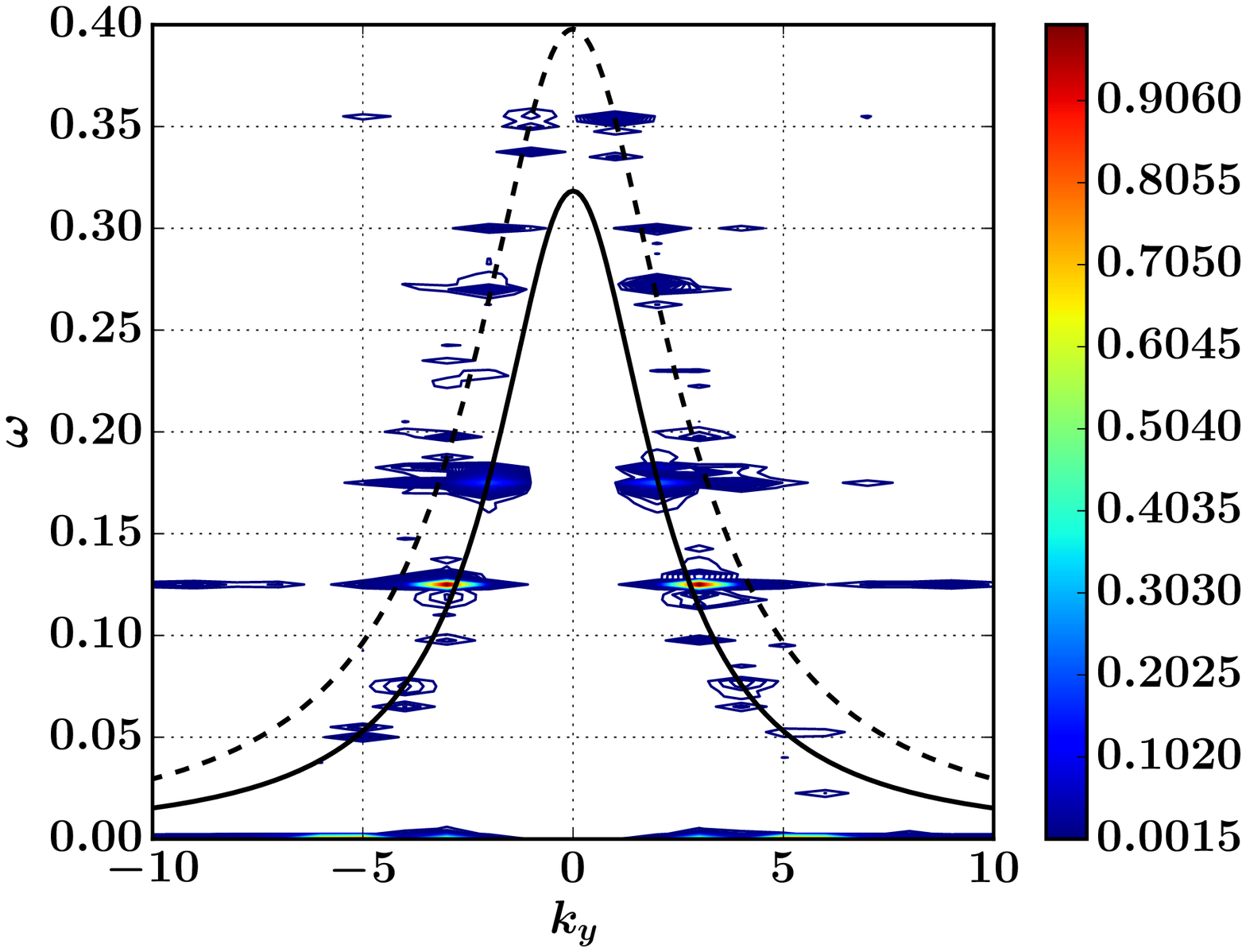}
\includegraphics[width=7cm,height=6cm]{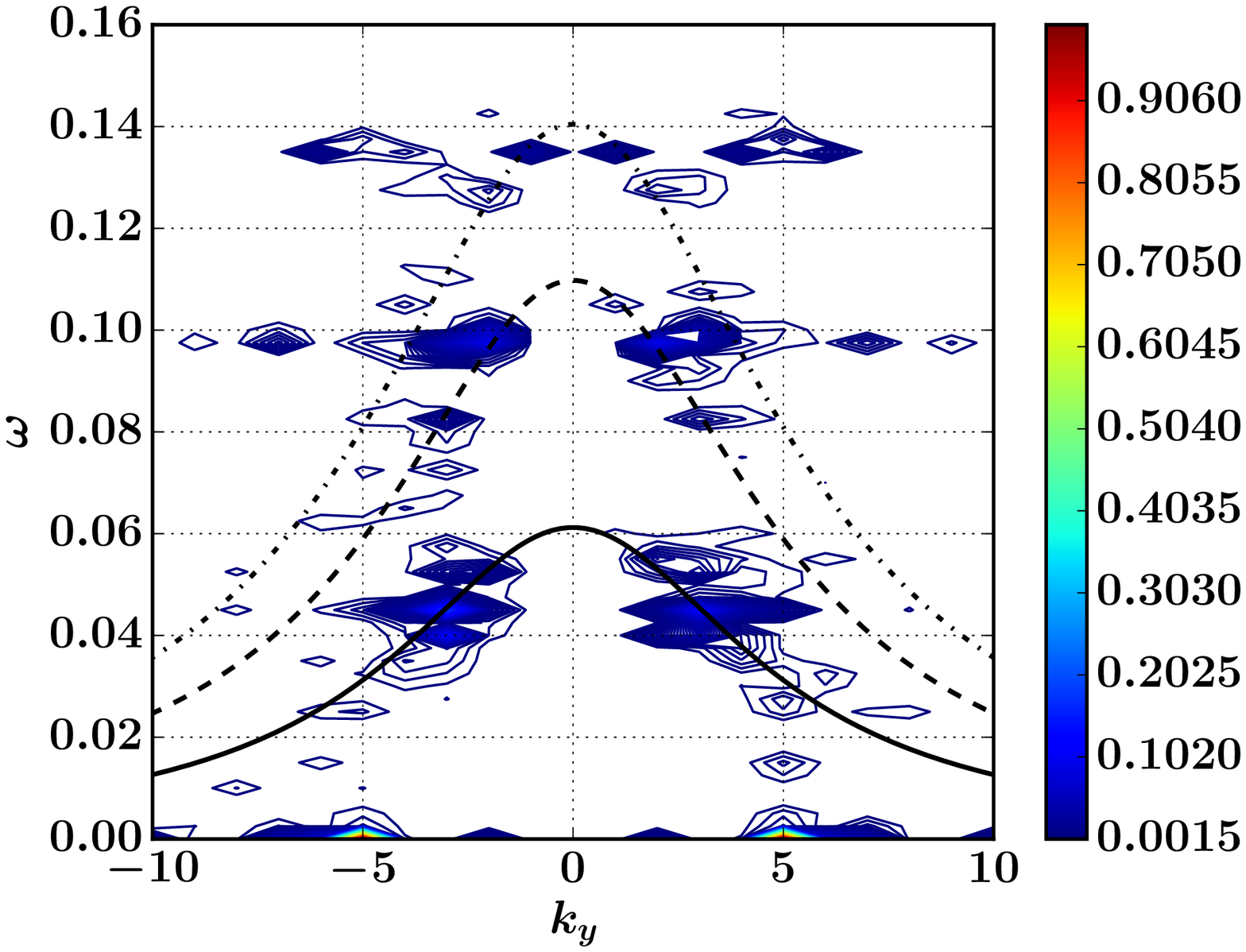}
\includegraphics[width=7cm,height=6cm]{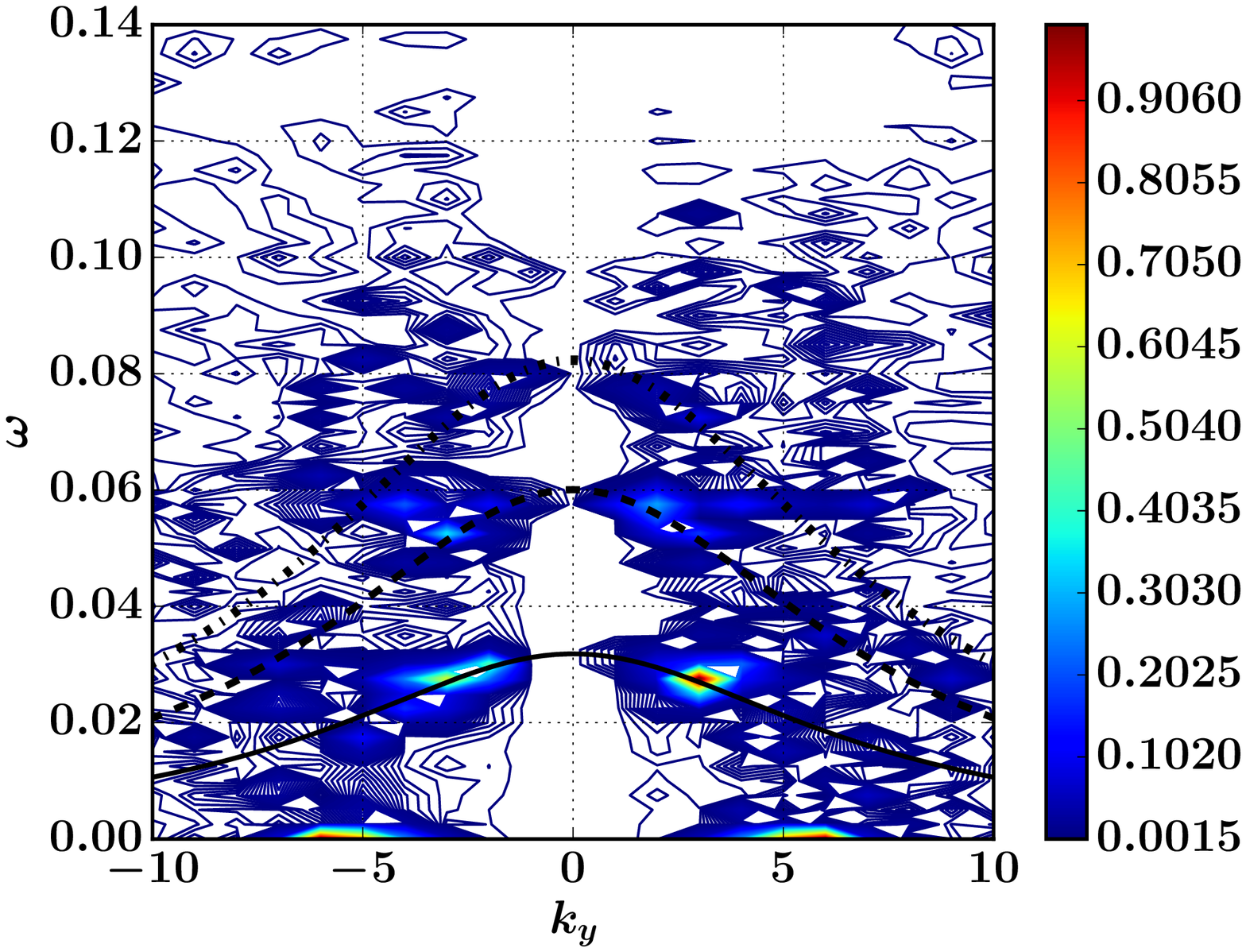}
\caption{\label{fky_jets} Frequency-meridional wavenumber plots constructed using 4000 sequential snapshots of the zonal velocity for $k_D=0,2,5$ and $7$ in 
the top left, top right, bottom left and bottom right panels, respectively. Solid,dashed and dash-dot curves are the Rossby dispersion
relations for $k_x=1,2$ and $3$.}
\end{figure}

Having restricted $k_y \approx 1/L_T$, we saw that much of the energy in the system was consistent with the Rossby dispersion curve and
was restricted to the first few zonal wavenumbers. Now, we reverse the situation, i.e., we fix $k_x$ to be small, as observed (at least when jets are
present), and examine the distribution of energy in frequency-meridional wavenumber diagrams. For $L_D=\infty, 1/2, 1/5$ and $1/7$, these plots are
shown in the four panels of Figure \ref{fky_jets}. As would be expected, there is no sign of a steady jet in this representation, in fact,
the peaks of
the spectrum (especially for $L_D=\infty, 1/2$ and $1/5$) align almost perfectly with the Rossby dispersion relation. Note that the plots are symmetric with respect to $\pm k_y$, i.e., 
the waves propagate in both, north and south
directions. In each case, as would be expected, the most prominent peak is at $k_y \approx 1/L_T$, but in addition we do see a significant amount of
energy at higher meridional wavenumbers.
Thus, taking both the frequency-zonal and frequency-meridional wavenumber diagrams into account, we identify that the modulation of jets is by
large-scale, low-frequency Rossby waves. 

\begin{figure}
\centering
\includegraphics[width=7cm,height=6cm]{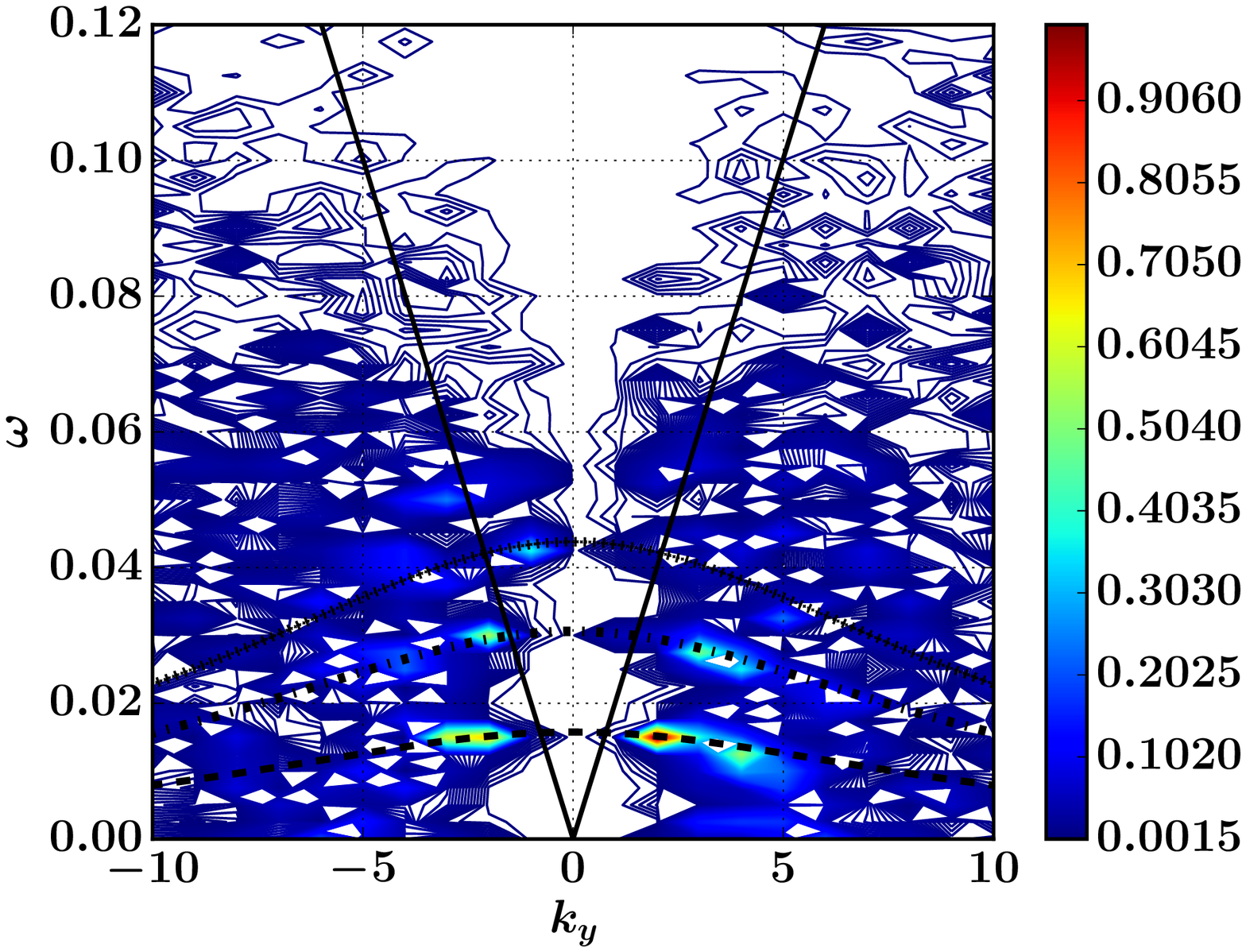}
\includegraphics[width=7cm,height=6cm]{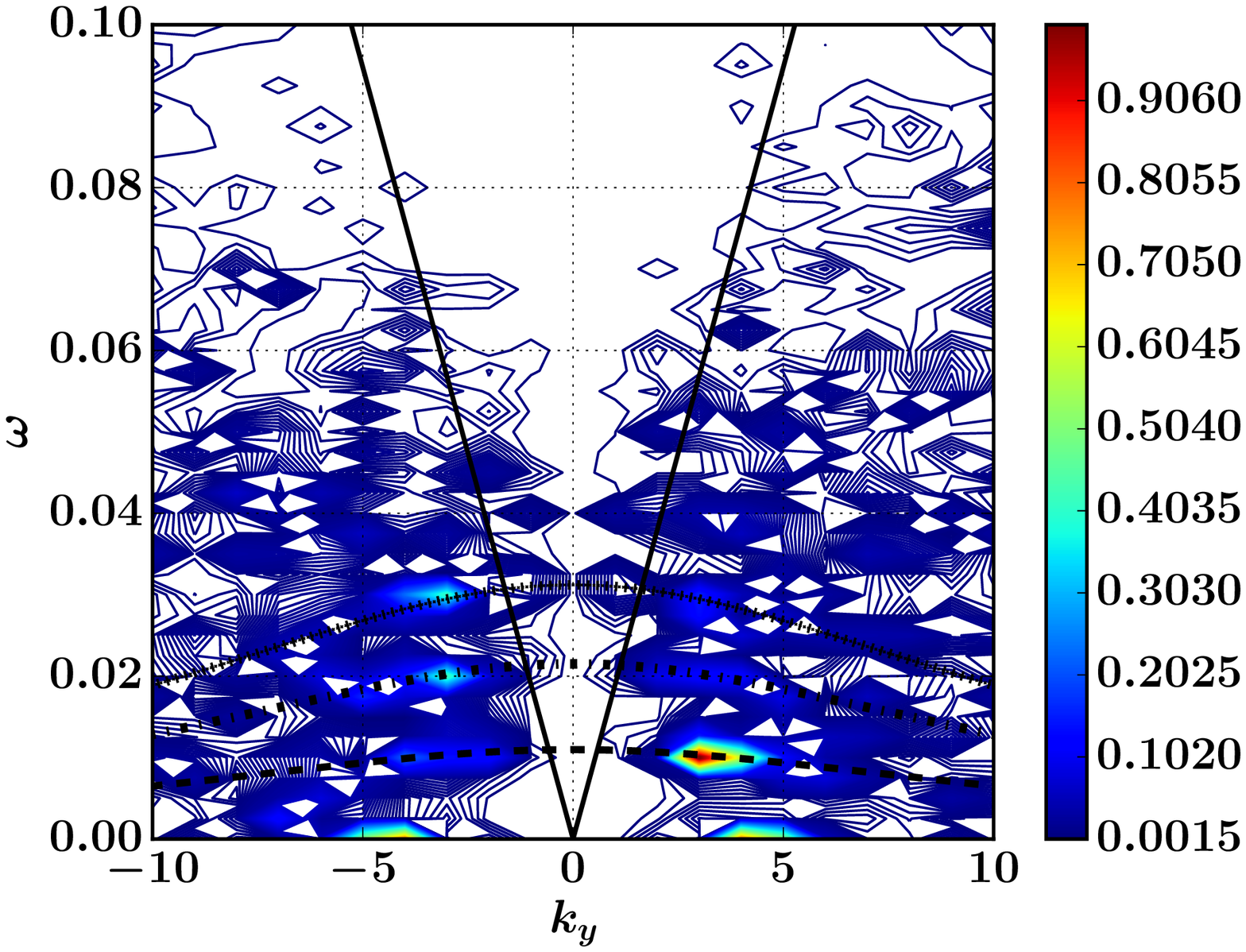}
\includegraphics[width=7cm,height=6cm]{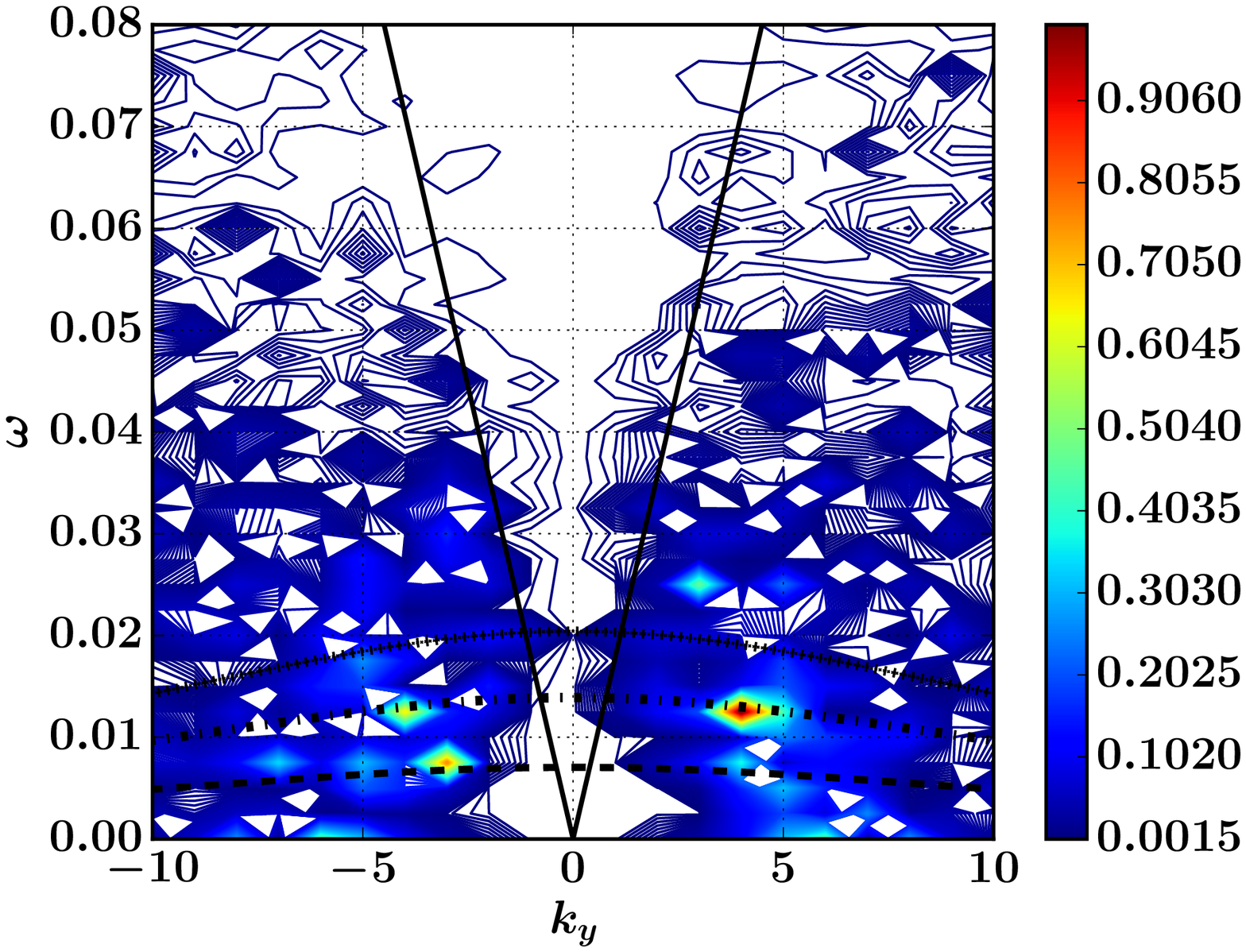}
\caption{\label{fky_nojets} Frequency-meridional wavenumber plots constructed using 4000 sequential snapshots of the zonal velocity for $k_D=10,12$ and $15$ in the top left, top right and bottom panels, respectively. Dashed, dash-dot and dotted curves are the Rossby dispersion
relations for $k_x=1,2$ and $3$ and solid line represents the turbulence frequencies, i.e., $Uk_y$. }
\end{figure}

By $L_D=1/7$ (the borderline case), we begin to see that the energy is distributed quite broadly in the frequency-meridional wavenumber domain.
This trend continues for
the cases when jets do not form, as is seen in Figure \ref{fky_nojets}. Even though, some peaks do align with higher $k_x$ Rossby waves,
the spectrum fills out
much of the region below the frequency associated with turbulence (i.e., $Uk_y$). Indeed, for a given $k_y$, the predominance of energy for
frequencies smaller than $Uk_y$ is consistent with the inverse transfer of energy in the frequency domain (Figure \ref{freqflux}).

\section{Discussion}

QG turbulence with a finite deformation radius has been examined numerically; our emphasis has been on an analysis of the problem in 
frequency and combined wavenumber-frequency domains. In line with recent work on the infinite deformation radius case, we have shown that
when jets emerge in the system, they are continuously modulated by low-frequency, large-scale Rossby waves. In particular, inspecting the 
scaling of the KE with frequency, we find some evidence of a power-law at high frequencies (associated with an inverse transfer of energy to low frequencies) 
that starts from the forcing frequency itself. Further, the scaling 
extends to lower frequencies for progressively smaller deformation scales; a feature consistent with the dispersion relation of Rossby waves.
But, whenever jets form, beyond the scaling regime, we note departures from a power-law in the form of isolated peaks in the spectrum.

In the presence of jets, the nature of the aforementioned peaks is investigated by means of frequency-meridional wavenumber and frequency-zonal wavenumber diagrams. 
Indeed, noting that
that the modified Rhines scale acts as a reasonable measure of the scale where energy piles up in the $k_y$ spectrum, we choose this as a meridional 
wavenumber. 
Remarkably, apart from the steady jet, 
the spectral peaks of energy in the system (on a frequency-zonal wavenumber plot) align themselves with large-scale, 
low-frequency westward propagating disturbances and are consistent with the
Rossby dispersion curves for $k_y \approx 1/L_T$. Reversing the situation, noting that the zonal scale of the modulating 
waves is restricted to the first few wavenumbers, we compare the corresponding dispersion curves with the spectral distribution of the 
energy on frequency-meridional wavenumber plots. Once again, along with anticipated peaks near the transition scale, we see that almost all of the
energy is closely confined to the Rossby dispersion curves. 

The range of parameters considered also allows us to examine non-jet forming cases. Here, the flow consists of small-scale, isotropic 
vortices that do not merge but retain their character for the duration of the simulations. The scaling of energy, and the change from KE to 
PE dominance with increasing length scale, as one crosses the deformation scale, is particularly evident in these cases. Indeed, here, the total energy
shows a prominent transition from steep (-11/3) to shallow (-5/3) scaling at the radius of deformation. In frequency space, the scaling of KE
is shallower than the jet-forming cases, and is closer to a -5/3 power-law that
extends to the lowest frequencies allowed by the data sampled. Interestingly, the PE falls off much more rapidly with frequency than the KE. 
Once again, like wavenumber space, the total energy shows a transition from a steep (PE dominated) to shallow (KE dominated) scaling at the ``deformation 
frequency". In terms of frequency-wavenumber diagrams, the energy is much more broadly distributed
than in the jet forming cases. 
In fact, there is a predominance of westward moving 
disturbances whose frequency is always less than the corresponding frequency associated with turbulence.

\vskip 0.5truecm
{\it Acknowledgements:}
The authors acknowledge 
financial support 
from the Divecha Centre for Climate Change at the Indian Institute of Science, Bangalore.


\begin{thebibliography}{99}

\bibitem{Rhines} P. B. Rhines, "Waves and turbulence of a beta plane," J. Fluid. Mech., 69, 417 (1975).

\bibitem{Vallis-book} G. K. Vallis, {\it Atmospheric and Oceanic Fluid Dynamics}, Cambrigde University Press (2005).

\bibitem{HH} G. Holloway and M. C. Hendershott, "Stochastic closure for nonlinear Rossby waves," J. Fluid. Mech., 82, 747 (1977).

\bibitem{Maltrud-Vallis} M. E. Maltrud and G. K. Vallis, "Energy spectra and coherent structures in forced two-dimensional and
beta-plane turbulence," J. Fluid. Mech., 228, 321 (1991).

\bibitem{Cheklov-etal} A. Cheklov, S. A. Orszag, S. Sukoriansky, B. Galperin and I. Staroselsky, "The effect of small-scale
forcing on large-scale structures in two-dimensional flows,"
Physica D, 98, 321 (1996).

\bibitem{Marcus} P. S. Marcus and C. Lee, "A model for eastward and westward jets in laboratory experiments and
planetary atmospheres,"
Phys. of Fluids, 10, 1474 (1998).

\bibitem{Danilov1} S. Danilov and V. M. Grynik, "Barotropic beta plane turbulence in a regime with strong zonal jets
revisited,"
J. Atmos. Sci., 61, 2283 (2004).

\bibitem{Gal-etal} B. Galperin, S. Sukoriansky, N. Dikovskaya, P. L. Read, Y. H. Yamazaki and R. Wordsworth, "Anisotropic turbulence and zonal jets in rotating flows with a $\beta$-effect,"
Nonlin. Processes Geophys., 13, 83 (2006).

\bibitem{Smith-qg} Y. Lee and L. M. Smith, "A mechanism for the formation of geophysical and planetary zonal flows," 
J. Fluid Mech., 576, 405 (2007).

\bibitem{SDG-2007} S. Sukoriansky, N. Dikovskaya and B. Galperin, "On the arrest of inverse energy cascade and the Rhines
scale," J. Atmos. Sci., 64, 3312 (2007).

\bibitem{McDrit} D. G. Dritschel and M. E. McIntyre, "Multiple jets as PV staircases: the Phillips effect and the resilience
of eddy-transport barriers," J. Atmos. Sci., 65, 855 (2008).

\bibitem{Scott1} R. K. Scott and D. G. Dritschel, "The structure of zonal jets in geostrophic turbulence," J. Fluid Mech.,
711, 576 (2012).

\bibitem{Hass} K. Hasselmann, "A criterion for nonlinear wave stability," J. Fluid Mech., 30, 737 (1967).

\bibitem{Cetal} C. Connaughton, B. Nadiga, S. Nazarenko and B. Quinn, "Modulational instability of Rossby and drift
waves and generation of zonal jets," J. Fluid Mech., 654, 207 (2010).

\bibitem{SrinYoung} K. Srinivasan and W. R. Young, "Zonostrophic instability," J. Atmos. Sci., 69, 1633 (2012).

\bibitem{BI-2014} N. A. Bakas and P. J. Ioannou, "A theory for the emergence of coherent structures in beta-plane
turbulence," J. Fluid Mech., 740, 312 (2014).

\bibitem{Con} C. Connaughton, S. Nazarenko and B. Quinn, "Rossby and drift wave turbulence and zonal flows: the
Charney-Hasegawa-Mima model and its extensions," arxiv.org/pdf/1407.1896 (2014).

\bibitem{Larichev-Mc} V. D. Larichev and J. C. McWilliams, "Weakly decaying turbulence in an equivalent-barotropic fluid,"
Phys. of Fluids, A3, 938 (1991).

\bibitem{Lipps} F. B. Lipps, "Stability of jets in a divergent barotropic fluid," J. Atmos. Sci., 20, 120 (1963).

\bibitem{Holloway} G. Holloway, "Effects of planetary wave propagation and finite depth on the predicability of atmospheres,"
J. Atmos. Sci., 40, 314 (1983).

\bibitem{Theiss} J. Theiss, "A generalized Rhines effect and storms on Jupiter," Geophys. Res. Lett., 33, L08809 (2006).

\bibitem{Penny} A. B. Penny, A. P. Showman and D. S. Choi, "Suppression of the Rhines effect and the location of
vortices on Saturn," J. Geophys. Res., 115, E02001 (2010).

\bibitem{Eden} C. Eden, "Eddy length scales in the north Atlantic ocean," Geophys. Res. Lett., 112, C06004 (2007).

\bibitem{Smith-2004} K. S. Smith, "A local model for planetary atmospheres forced by small-scale convection," J. Atmos. Sci.,
51, 1420 (2004).

\bibitem{KOY-1995} N. Kukharkin, S. A. Orszag and V. Yakhot, "Quasicrystallization of vortices in drift-wave turbulence,"
Phys. Rev. Lett., 75, 2486 (1995).

\bibitem{ISW-2002} T. Iwayama, T. G. Shepherd and T. Watanabe, "An ideal form of decaying two-dimensional turbulence,"
J. Fluid Mech., 456, 183 (2002).

\bibitem{Tran-Bowman} C. V. Tran and J. C. Bowman,
"Energy budgets in Charney-Hasegawa-Mima and surface quasi-
geostrophic turbulence," Phys. Rev. E, 68, 036304 (2003).

\bibitem{Tran-Drit} C. V. Tran and D. G. Dritschel, "Impeded inverse energy transfer in the Charney-Hasegawa-Mima model
of quasi-geostrophic flows," J. Fluid Mech., 551, 435 (2006).

\bibitem{Scott2} R. K. Scott and D. G. Dritschel, "Halting scale and energy equilibration in two-dimensional quasi-
geostrophic turbulence," J. Fluid Mech., 721, R4, (2013).

\bibitem{SS-2009} J. Sukhatme and L. M. Smith, "Local and nonlocal dispersive turbulence," Phys. of Fluids, 21, 056603
(2009).

\bibitem{Okuno} A. Okuno and A. Masuda, "Effect of horizontal divergence on the geostrophic turbulence on a beta-plane:
Suppression of the Rhines effect," Phys. of Fluids, 15, 53 (2003).

\bibitem{Suko-PRL} S. Sukoriansky, N. Dikovskaya and B. Galperin, "Nonlinear waves in zonostrophic turbulence," Phys.
Rev. Lett., 101, 178501 (2008).

\bibitem{Afan} Y. Zhang and Y. D. Afanasyev, "Beta-plane turbulence: Experiments with altimetry," Phys. of Fluids,
26, 026602–1 (2014).

\bibitem{Arbic-2012} B. K. Arbic, R. B. Scott, G. R. Flierl, A. J. Morten, J. G. Richman and J. F. Shriver, "Nonlinear
cascades of surface oceanic geostrophic kinetic energy in the frequency domain," J. Phys. Oceanography,
42, 1577 (2012).

\bibitem{BK1} P. Berloff and I. Kamenkovich, "On spectral analysis of mesoscale eddies. Part I: Linear analysis," J.
Phys. Oceanography, 43, 2505 (2013).

\bibitem{BK2} P. Berloff and I. Kamenkovich, "On spectral analysis of mesoscale eddies. Part II: Nonlinear analysis," J.
Phys. Oceanography, 43, 2528 (2013).

\bibitem{Danilov2001} S. Danilov and D. Gurarie, "Forced two-dimensional turbulence in spectral and physical space," Phys.
Rev. E, 63, 061208 (2001).

\bibitem{WK-1999} M. Wheeler and G. H. Kiladis, "Convectively coupled equatorial waves: Analysis of clouds and temperature
in the wavenumber-frequency domain," J. Atmos. Sci, 56, 374 (1999).

\bibitem{Ten} H. Tennekes and J. L. Lumley, {\it A First Course in Turbulence},  MIT Press (1972).

\bibitem{Landau} L. D. Landau and E. M. Lifshitz, {\it Fluid Mechanics}, Pergamon Press (1987).

\bibitem{FW-2010} R. Ferrari and C. Wunsch, "The distribution of eddy kinetic and potential energies in the global ocean,"
Tellus, 62A, 92 (2010).


\end{thebibliography}
\end{document}